\documentclass[preprint,review,12pt]{elsarticle}

\pdfoutput=1
\usepackage[utf8]{inputenc}
\usepackage{amsmath}
\usepackage{amsfonts}
\usepackage{amssymb}
\usepackage[pdftex]{hyperref}
\usepackage{graphicx}
\usepackage{subfigure}
\usepackage{algorithmic}
\usepackage{algorithm}
\usepackage{color}
\usepackage{bm}

\biboptions{sort&compress}

\definecolor{darkgreen}{rgb}{0,0.6,0.2}

\journal{Comm. in Nonlinear Science and Numerical Sim}

\makeindex

\begin{document}
\begin{frontmatter}

\title{Right-side-stretched multifractal spectra indicate small-worldness in networks}

\author[address1]{Pawe\l{} O\'{s}wi\c{e}cimka}
\ead{pawel.oswiecimka@ifj.edu.pl}
\author[address2]{Lorenzo Livi\corref{cor1}}
\cortext[cor1]{Corresponding Author}
\ead{l.livi@exeter.ac.uk}
\ead[url]{https://sites.google.com/site/lorenzlivi/}
\author[address1,address3]{Stanis\l{}aw Dro\.{z}d\.{z}}
\ead{stanislaw.drozdz@ifj.edu.pl}

\address[address1]{Complex Systems Theory Department, Institute of Nuclear Physics, Polish Academy of Sciences, PL-31-342 Krak\'{o}w, Poland}
\address[address2]{Department of Computer Science, College of Engineering, Mathematics and Physical Sciences, University of Exeter, Exeter EX4 4QF, UK}
\address[address3]{Faculty of Physics, Mathematics and Computer Science, Cracow University of Technology, PL-31-155 Krak\'{o}w, Poland}

\begin{abstract}
Complex network formalism allows to explain the behavior of systems composed by interacting units. Several prototypical network models have been proposed thus far. The small-world model has been introduced to mimic two important features observed in real-world systems: i) local clustering and ii) the possibility to move across a network by means of long-range links that significantly reduce the characteristic path length.
A natural question would be whether there exist several ``types'' of small-world architectures, giving rise to a continuum of models with properties (partially) shared with other models belonging to different network families.
Here, we take advantage of the interplay between network theory and time series analysis and propose to investigate small-world signatures in complex networks by analyzing multifractal characteristics of time series generated from such networks.
In particular, we suggest that the degree of right-sided asymmetry of multifractal spectra is linked with the degree of small-worldness present in networks.
This claim is supported by numerical simulations performed on several parametric models, including prototypical small-world networks, scale-free, fractal and also real-world networks describing protein molecules. Our results also indicate that right-sided asymmetry emerges with the presence of the following topological properties: low edge density, low average shortest path, and high clustering coefficient.
\end{abstract}
\end{frontmatter}

\section{Introduction}

Nonlinear time series analysis \cite{bradley2015nonlinear} and network theory \cite{bianconi2015interdisciplinary} have become standard frameworks for analyzing complex systems.
Interesting hybridizations of the two frameworks are becoming popular as well, providing the possibility to study system dynamics observed as time series but analyzed in terms of topological features of associated complex network \cite{budroni2016scale,0295-5075-116-5-50001,wang2016data,weng2017memory}; analogously, complex networks can be studied in terms of time series analysis. For instance, recent works indicate the possibility to characterize network structures in terms of scaling and related (multi)fractal properties of suitably generated time series \cite{nicosia2013characteristic,mixbionets2,weng2014time,mfcrosscorr_nets}.

A number of complex network models have been proposed in the literature \cite{newman2010networks,lambiotte2007majority}.
Among the many, the so-called ``small-world'' networks have had substantial impact on the understanding of a wide range of complex natural and technological systems \cite{amaral2000classes}, such as brain \cite{bassett2016small} and metabolic networks \cite{fell2000small}, and smart electrical grids \cite{Pagani20132688}.
An interesting practical issue with theoretical implications is whether real-world complex systems can be completely associated with prototypical networks within, for instance, the small-world family or they are actually better described as hybrids of different models. The idea of a ``network morphology'' \cite{avena2015network} comes handy here, conceptualizing parametric network models on a configuration space, where different regions of such a space are occupied by networks with different theoretical/practical properties.
Efforts in this direction include two recent works of particular interest. The first one \cite{gallos2012small} studies the necessary architectural trade-off needed to obtain modular networks (having fractal intra-module topology) with small-world features found in brain networks; the second one provides a theoretical model describing phase transitions between small-world and fractal network models \cite{rozenfeld2010small}, properties that cannot otherwise co-exist in a single complex network.

Considering the recent data deluge and relaxing the assumption of a general clear-cut between network models, it seems reasonable to design methods that can provide a quantitative measure expressing the membership of a given experimental network to a network family, e.g., the small-world one. To this end, \citet{humphries2008network} suggested a measure of ``small-worldness'', hence providing a way to quantify how much a given network is actually a small-world network.

In this paper, we elaborate over the dualism between complex networks and time series analysis and show that fractal analysis of time series can be used to judge over the degree of small-worldness present in networks.
Specifically, we frame our contribution between random walks \cite{burioni2005random,PhysRevE.88.012817} and multifractal analysis of time series \cite{mukli2015multifractal,grech2016alternative,kwapien2012physical}. Random walks are used to generate time series of suitable vertex properties, which are successively analyzed within the multifractal analysis framework.
The novelty of our contribution can be summarized as follows:
\begin{itemize}

 \item We propose to use the degree of right-sided asymmetry of multifractal spectra \cite{drozdz2015detecting}, i.e., the degree to which spectra are stretched on the right-hand side, estimated for time series generated from complex networks as a signature of small-worldness in the corresponding networks. This claim is supported by experimental results on Watts-Strogatz \cite{watts1998collective}, Dorogovtsev-Goltsev-Mendes \cite{dorogovtsev2002pseudofractal}, Song-Havlin-Makse \cite{song2006origins} network models and on real data describing protein contact networks \cite{protgen1};
 
 \item The proposed criterion based on right-sided spectrum asymmetry can be applied also to networks where local clustering coefficient cannot be defined (e.g., in tree-like networks). For instance, on Song-Havlin-Makse networks clustering coefficient cannot be computed. Hence, conventional measures of small-worldness cannot be used. On the other hand, we show that the degree of right-sided asymmetry can be used also in these cases as a reliable indicator of network small-worldness;

 \item We argue that low edge density, low average shortest path, and high clustering coefficient are topological properties linked with the emergence of right-sided spectrum asymmetry of at least one of the three vertex observables taken into account. This finding is connected with the results shown in Ref. \cite{humphries2008network}, where the authors suggest a measure of small-worldness and show that when edge density is increased, Watts-Strogatz networks become indistinguishable from Erd\"{o}s-R\'{e}nyi graphs, hence losing their characteristic small-world signature.

\end{itemize}

The remainder of the paper is structured as follows.
Section \ref{sec:methods} introduces the methods used for the analysis.
In Section \ref{sec:experiments}, we present and discuss the results of our analysis performed on various networks.
Conclusions and future directions follow in Section \ref{sec:conclusions}

\section{Methods}
\label{sec:methods}

\subsection{Random walks on graphs and related time series of vertex properties}
\label{sec:rw}

Let $G=(\mathcal{V}, \mathcal{E})$ be an undirected graph, with $\mathcal{V}, n=|\mathcal{V}|,$ and $\mathcal{E}$ denoting the sets of vertices and edges, respectively.
A Markovian random walk in a graph \cite{masuda2016random} is a first-order Markov chain on the graph vertices.
Transitions among vertices are dictated by the transition matrix,
\begin{equation}
\label{eq:transition_matrix}
[0, 1]^{n\times n}\ni\mathbf{T}=\mathbf{D}^{-1}\mathbf{A},
\end{equation}
where $\mathbf{A}$ is the adjacency matrix of $G$ and $\mathbf{D}$ is a diagonal matrix of vertex degrees, i.e., $D_{ii}=\mathrm{deg}(v_i)=\sum_{j=1}^{n} A_{ij}$.

Let $\mathbf{p}_{0}$ be an initial distribution for the chain. The probability of the states at time $t>0$ can be computed by the following recursion: $\mathbf{p}_{t}=\mathbf{p}_{t-1}\mathbf{T}=\mathbf{p}_{0}\mathbf{T}^{t}$.
The stationary distribution of the chain, $\boldsymbol\pi$, is a probability vector satisfying $\boldsymbol\pi=\boldsymbol\pi\mathbf{T}$.
If the graph is undirected and non-bipartite, then it always possesses a (unique) stationary distribution that can be easily computed from the degree distribution, $\pi_i = \mathrm{deg}(v_i)/(2|\mathcal{E}|)$.
A stationary random walk is hence completely described by $\boldsymbol\pi$.

Let us define a time-homogeneous vertex property map as $M_{P}: \mathcal{V}\rightarrow\mathcal{O}_{P}$, where $\mathcal{O}_{P}$ is the domain of vertex property $P$, such as degree, closeness centrality or other, user-defined vertex properties.
By performing a (stationary) random walk on $G$, we generate a sequence of vertices, $(v_{i_{0}}, v_{i_{1}}, ..., v_{i_{T}})$, which are visited over a finite yet sufficiently large time span $T\gg 1$.
It is possible to associate to such a random walk a sequence of observables by emitting at each time instant $t$ the corresponding vertex property given by $O_t=M_{P}(v_{i_{t}})$, where $v_{i_{t}}$ is the i$th$ vertex of $G$ that is visited at time $t$.
This process generates a sequence of vertex properties, $\mathcal{S}_{P}=(O_0, O_{1}, ..., O_{T})$.
It is worth noting that a first-order Markov process is by definition a memory-less process. Nonetheless, as first observed in \cite{mixbionets2,nicosia2013characteristic,weng2014time}, time series $\mathcal{S}_{P}$ of vertex properties emitted by such a process might posses a complex organization characterized by a non-trivial correlation structure, which can be quantified by means of the multifractal analysis framework. As in \cite{mfcrosscorr_nets}, here we consider the following properties: vertex degree (VD), clustering coefficient (VCL), and closeness centrality (VCC).

\subsection{Scaling of fluctuations and multifractal analysis}
\label{sec:fluctuations}

Self-similarity of time series can be investigated by focusing on the scaling properties of the fluctuations, $F(s)$, expressed as a function of the length-scale $s>0$ \cite{havlin1999scaling}. Self-similar signals posses power-law scaling of the fluctuations, i.e. $F(s)\sim s^H$, where $H$ is called Hurst exponent. When $H=1/2$ only short-term correlations are present in the signal; $H>1/2$ indicates persistency of the underlying process, i.e., the presence of long-term correlations, while $H<1/2$ points out the presence of anti-correlations \cite{holl2015relationship}.
Several different methods have been proposed to analyze the fluctuations of the stationary component of a time series \cite{riley2012tutorial,fernandez2013measuring,wendt2007multifractality}.
The Detrended Fluctuation Analysis (DFA) procedure is a state-of-the-art method conceived for this purpose \cite{kiyono2016nonlinear}.
Its generalization, called Multifractal Detrended Fluctuation Analysis (MF-DFA) \cite{kantelhardt2002multifractal,PhysRevE.74.016103}, accounts also for multi-scaling, allowing a multi-level characterization of time series.

MF-DFA operates by analyzing the profile of a time series, which is then divided in $N_s$ non-overlapping segments of equal length $s$.
In order to account for non-stationarities due to trends, a polynomial function is fitted on (and then removed from) each of the $2N_s$ segments.
The procedure then computes the local variance, denoted as $F_2(\nu,s)$, for each segment $\nu = 1,\dotsc,2N_s$.
The $q$th-order moment of the local variance is given by $F_q(s) = \{ 1/2N_s \sum_{\nu=1}^{2N_s} \left[ F_2(\nu,s)\right]^{q/2} \}^{1/q}$, where $q \in \mathbb{R}$ operates as a magnification factor (to analyze the time series at different resolutions).
In order to check if the scaling of the variance follows a power law, i.e., $F_q(s) \sim s^{h(q)}$, the last steps are repeated for several scales $s$ of increasing length.
The generalized Hurst exponent, denoted as $h(q)$, characterizes the $q$-dependency of the fluctuation scaling.
A time series is called monofractal if $h(q)$ does not depend on $q$. On the other hand, when the dependency is strong enough, the time series is said multifractal.

The multifractal spectrum (MFS), denote as $f(\cdot)$, describes the multifractal properties of a time series and is given by
\begin{equation}
\label{eq:mutifractal_spectrum}
f(\alpha) = q(\alpha - h(q)) + 1,
\end{equation}
where $\alpha = h(q) + qh'(q)$ is called H\"older exponent.
It is worth noting that $f(\alpha)$ corresponds to the Hausdorff dimension of a subset of the input data where the H\"older exponent is equal to $\alpha$.
The multifractal spectrum (\ref{eq:mutifractal_spectrum}) encodes important information regarding the sensitivity to fluctuations with high/low magnitudes, analyzed by using positive/negative $q$ values.
The width of the support of $f(\cdot)$ is defined as
\begin{equation}
\label{eq:mfs_width}
\Delta\alpha=\alpha(q_{\textrm{min}})-\alpha(q_{\textrm{max}}), 
\end{equation}
where $q_{\textrm{min}}$ and $q_{\textrm{max}}$ denote the lower and upper end points of $q$, respectively.
The width of the spectrum offers an important quantitative complexity measure: the larger the width, the higher the complexity of the time series.
In addition, the degree of asymmetry $A_{\alpha}$ of MFS (\ref{eq:mutifractal_spectrum}) is another important feature to be taken into account \cite{drozdz2015detecting},
\begin{equation}
\label{eq:mfs_asymmetry}
A_{\alpha} = \frac{\Delta\alpha_{\text{L}}-\Delta\alpha_{\text{R}}}{\Delta\alpha_{\text{L}}+\Delta\alpha_{\text{R}}},
\end{equation}
where $\Delta\alpha_{\text{L}}$ and $\Delta\alpha_{\text{R}}$ denote the width of the left and right part of $f(\alpha)$, respectively.
A negative value for $A_{\alpha}$ implies right-sided asymmetry (i.e., spectra are stretched on the right-hand side), highlighting a stronger multifractality on smaller fluctuations. Conversely, left-sided spectra denote a higher heterogeneity for large fluctuations \cite{drozdz2015detecting}.

\subsection{Quantitative measures of small-worldness}
\label{sec:smallworldness}

A quantitative measure of small-worldness is defined in \cite{humphries2008network} as follows:
\begin{equation}
\label{eq:smallworldness}
S=\frac{\gamma^{\mathrm{ws}}_{g}}{\lambda_g},
\end{equation}
where 
\begin{align}
\label{eq:clustering_ratio}
\gamma^{\mathrm{ws}}_{g}=\frac{C_{g}^{\mathrm{ws}}}{C_{\mathrm{rand}}^{\mathrm{ws}}}, \\
\label{eq:asp_ratio}
\lambda_g=\frac{L_g}{L_{\mathrm{rand}}}. 
\end{align}

In Eq. \ref{eq:clustering_ratio}, $C_{g}^{\mathrm{ws}}$ is the average clustering coefficient of a given WS network $g$,
\begin{align}
\label{eq:avg_clustering_coeff}
C_{g}^{\mathrm{ws}} &= \langle c_{i}^{\mathrm{ws}} \rangle_i, \\
\label{eq:clustering_coeff}
c_{i}^{\mathrm{ws}} &= \frac{2m_i}{k_i(k_i-1)},
\end{align}
where $c_{i}^{\mathrm{ws}}$ is the clustering coefficient of $i$th vertex, $m_i$ and $k_i$ denote, respectively, the number of edges between the direct neighbors of vertex $i$ and its degree.
Analogously, $C_{\mathrm{rand}}^{\mathrm{ws}}$ denotes the average clustering coefficient of the corresponding Erd\"{o}s-R\'{e}nyi (ER) network with the same edge density, which is defined as:
\begin{equation}
\label{eq:edge_density}
\zeta=2m/(n(n-1)),
\end{equation}
where $m$ and $n$ are the number of edges and vertices, respectively.
Eq. \ref{eq:asp_ratio} is defined in a similar way but using average shortest path (ASP) instead of clustering coefficient.

As discussed in \cite{humphries2008network}, the categorical definition of small-world network given above implies $\lambda_g\geq 1$ and $\gamma^{\mathrm{ws}}_{g}\gg 1$. This fact imposes $S>1$ for small-world networks: the larger the value of $S$, the higher the degree of small-worldness.

\section{Results}
\label{sec:experiments}

The experimental results presented in this paper have been obtained by generating random walks with $T=10^6$ time-steps; results are always intended as averages of ten different random realizations of such random walks.
A recent study \cite{nicosia2013characteristic} points to 
the differences between short- and long-term correlations in the considered vertex observables. In particular, the data are strongly autocorrelated on short time-scales, whereas they can be considered as independent on large time-scale hence giving rise to a pronounced cross-over in the fluctuation functions.
Therefore, in this paper we concentrate mainly on short-range correlations (i.e., scales $s<800$) as more important from the perspective of the network structure here taken into account.

\subsection{Watts-Strogatz model}

Here we discuss the outcome of the analysis conducted on Watts-Strogatz (WS) models \cite{watts1998collective}.
There are two main parameters affecting the topology of WS networks: the number $k$ of nearest neighbors in the initial ring-like topology and the probability $p$ of rewiring an edge to a uniformly chosen pair of vertices.
In our experiments, $k$ assumes values in $\{4, 6, ..., 24\}$ and $p$ in $[0, 1]$. Notably, for each value of $k$, we select ten network configurations by increasing $p$ with a step of 0.1. When $p=1$, all edges are randomly rewired, hence heavily affecting the resulting WS network topology. WS networks are always composed by $10^3$ vertices. 

\begin{figure}[ht!]
\centering

\subfigure[Scaling of fluctuations for $p=0.1$.]{
\includegraphics[scale=0.3,keepaspectratio=true]{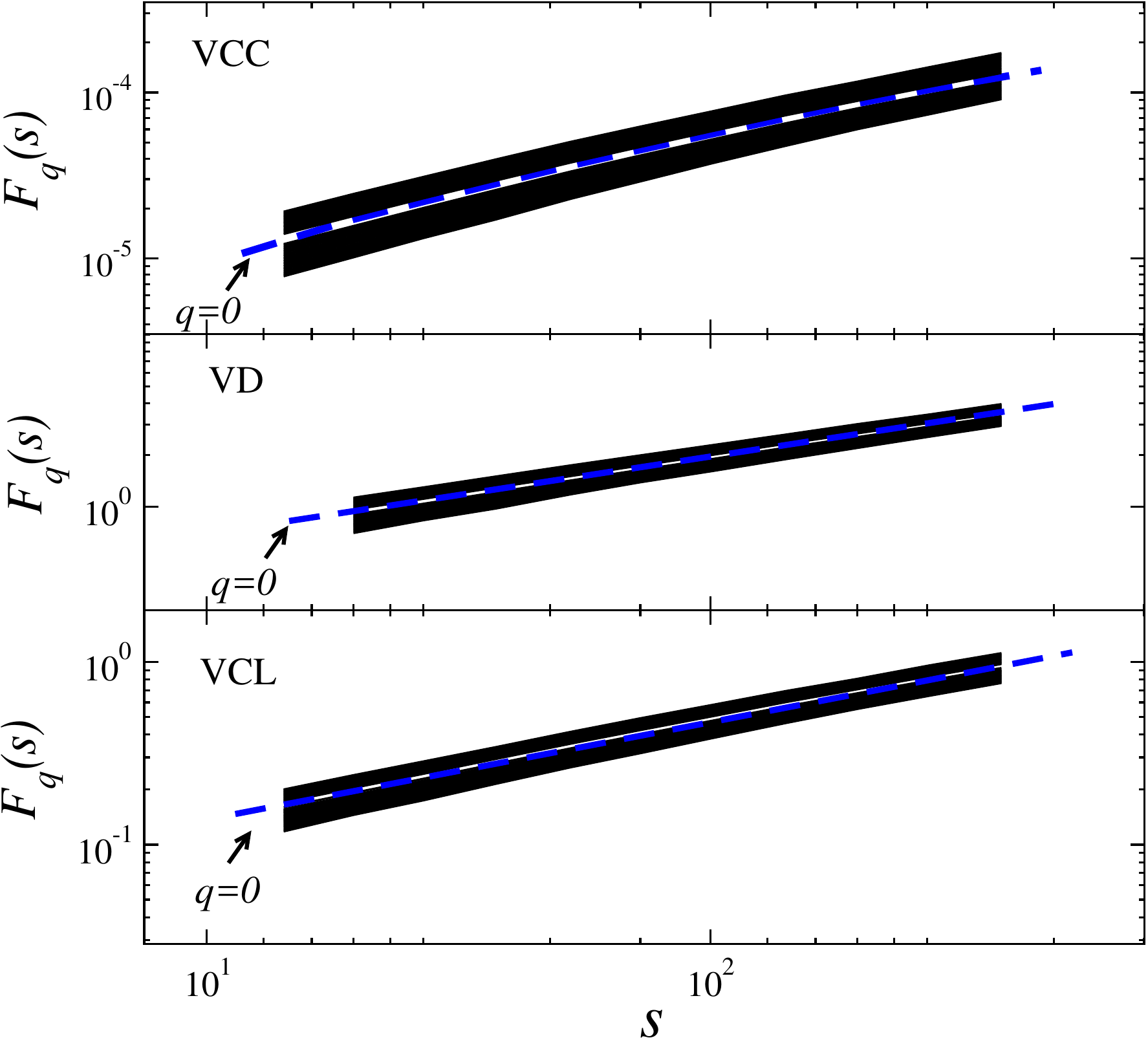}
\label{fig:WS11_fq}}
~
\subfigure[Multifractal spectra for $p=0.1$.]{
\includegraphics[scale=0.3,keepaspectratio=true]{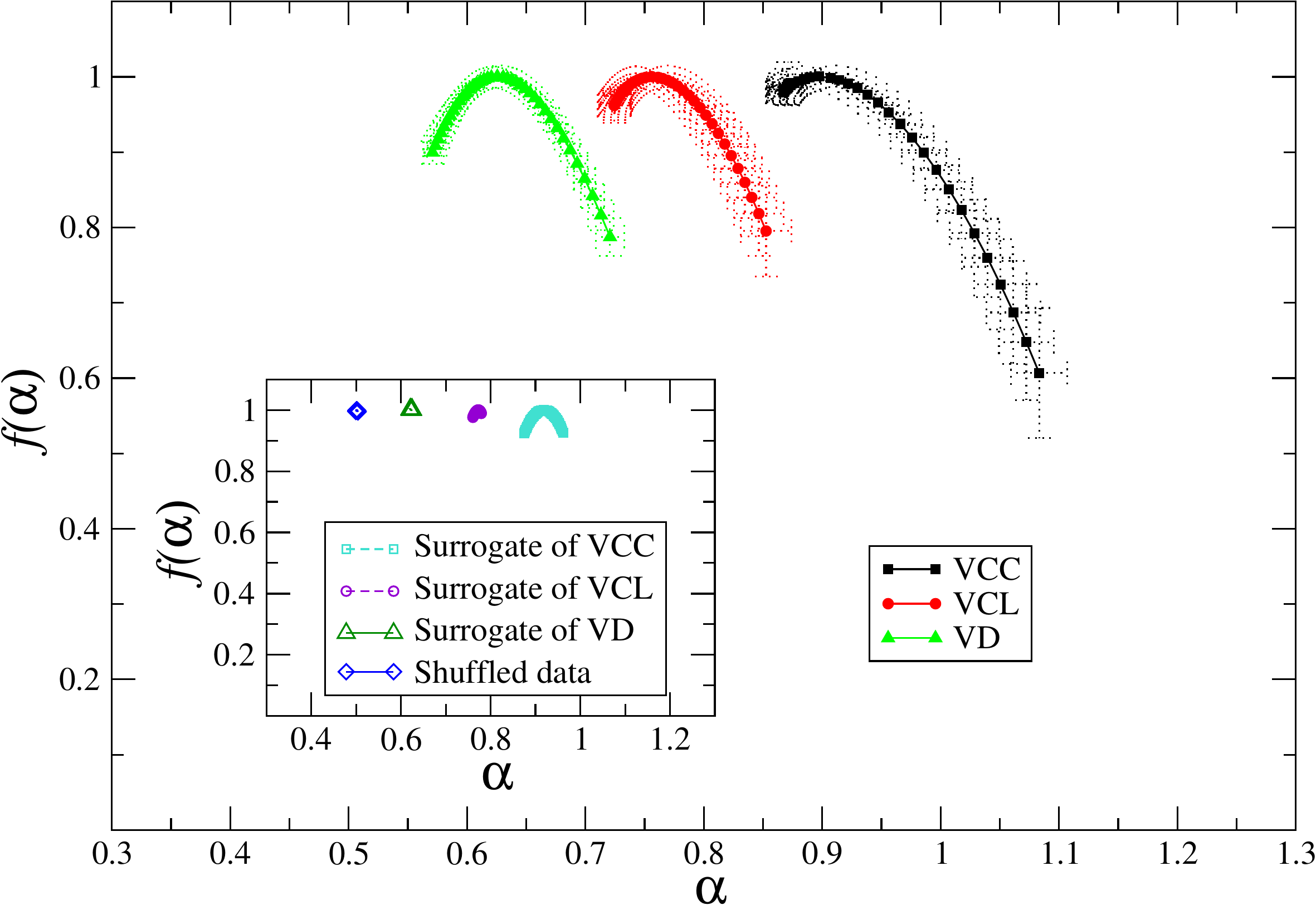}
\label{fig:WS11_falpha}}

\subfigure[Scaling of fluctuations for $p=0.3$.]{
\includegraphics[scale=0.3,keepaspectratio=true]{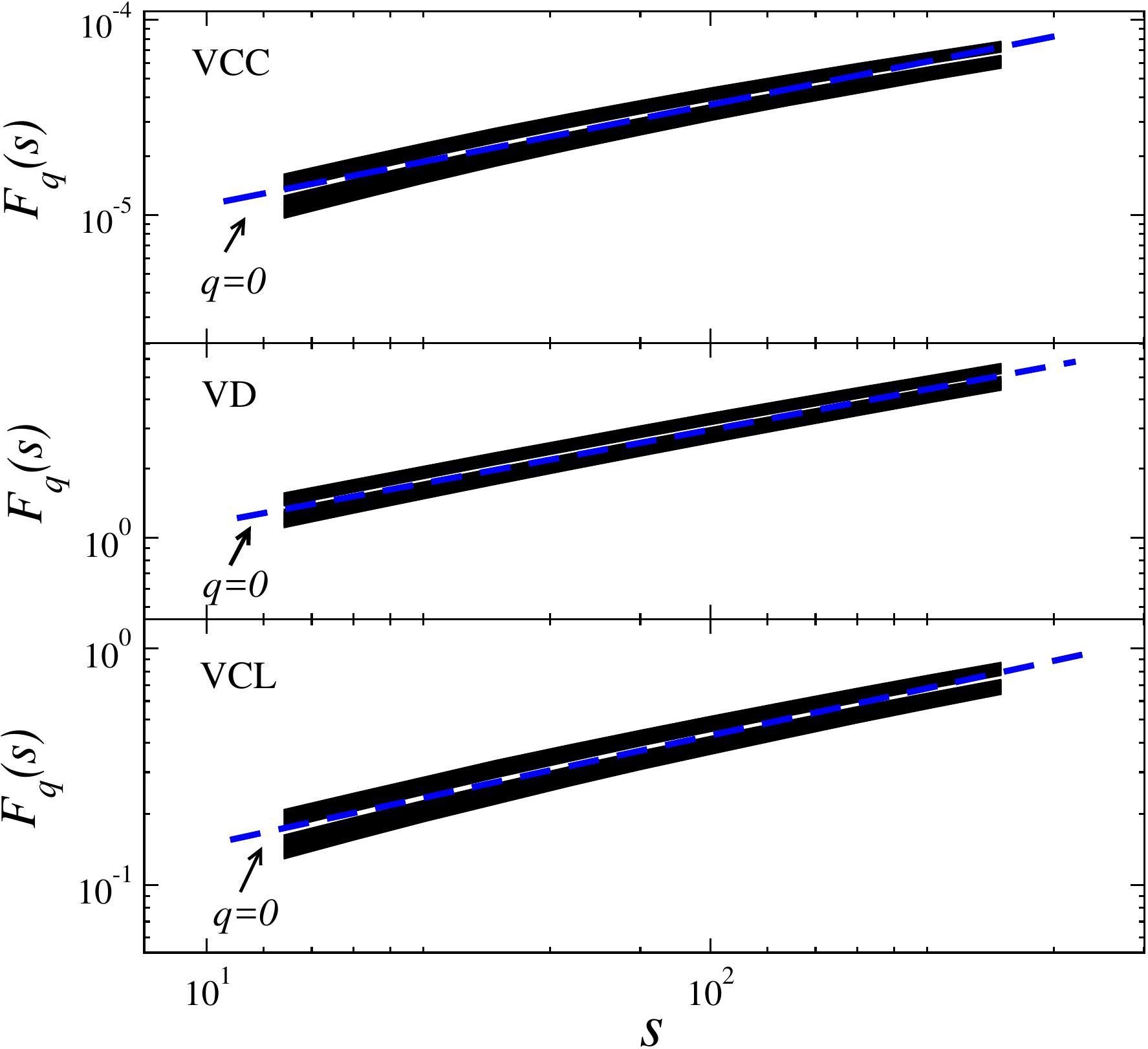}
\label{fig:WS13_fq}}
~
\subfigure[Multifractal spectra for $p=0.3$.]{
\includegraphics[scale=0.3,keepaspectratio=true]{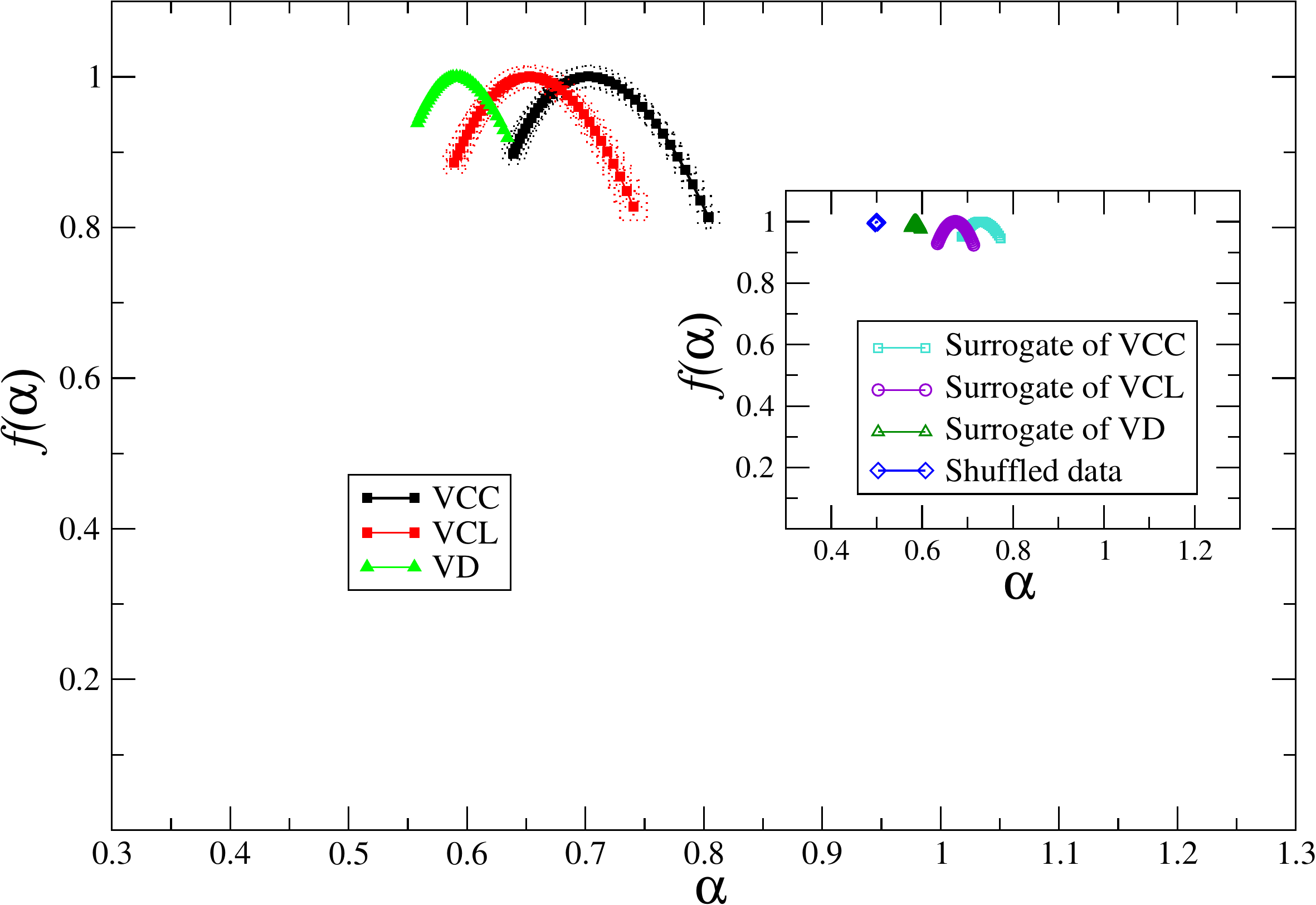}
\label{fig:WS13_falpha}}

\subfigure[Scaling of fluctuations for $p=0.9$.]{
\includegraphics[scale=0.3,keepaspectratio=true]{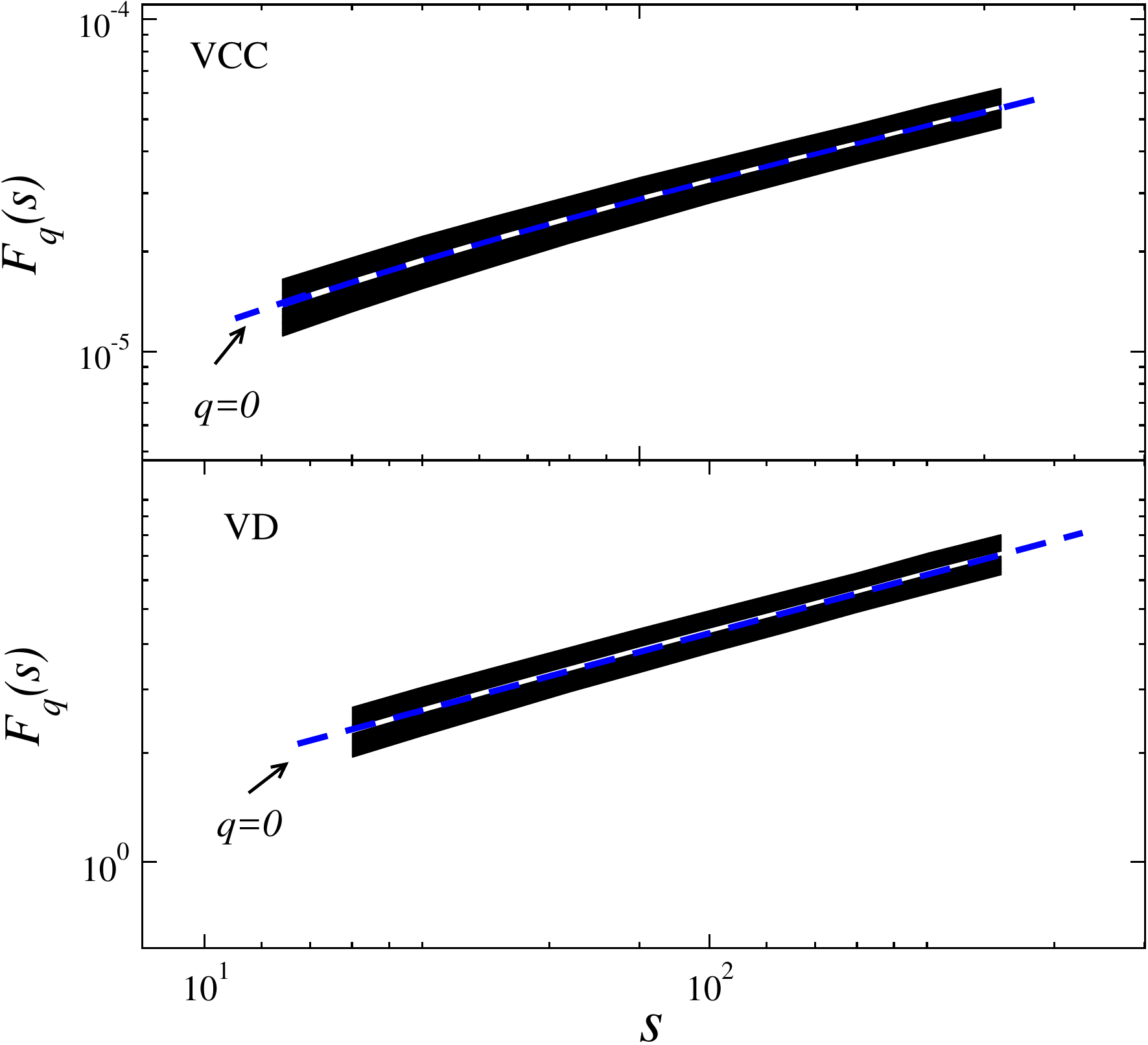}
\label{fig:WS19_fq}}
~
\subfigure[Multifractal spectra for $p=0.9$.]{
\includegraphics[scale=0.3,keepaspectratio=true]{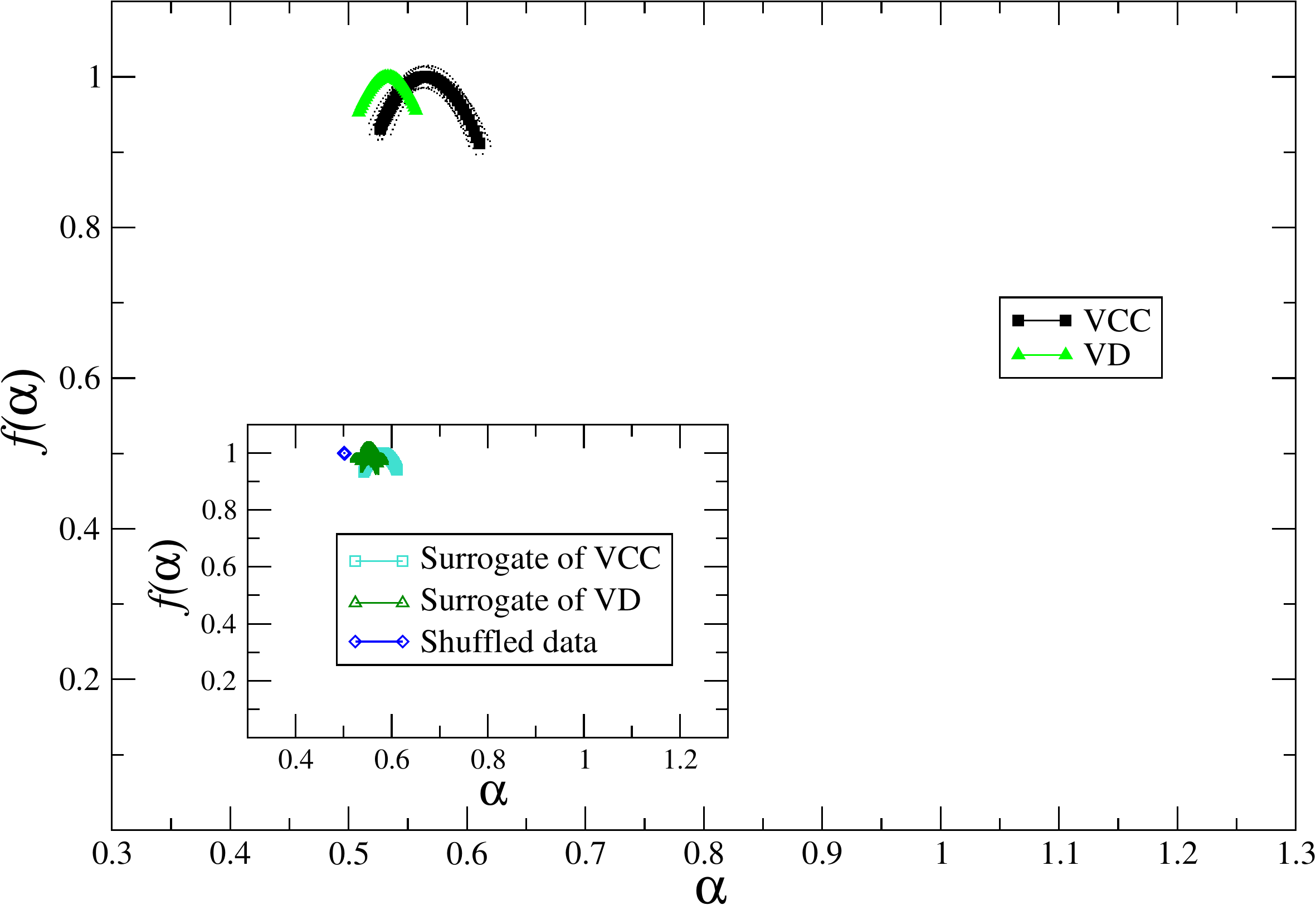}
\label{fig:WS19_falpha}}

\caption{Results for WS networks with $k=6$ fixed and rewiring probability $p=0.1, 0.3$, and $p=0.9$. Multifractality and right-sided asymmetry of MFS are systematically lost when increasing $p$. Results for VCL time series and $p=0.9$ are not shown due to the computational artefacts mentioned in the text.}
\label{fig:WS11_13_19}
\end{figure}

Fig. \ref{fig:WS11_13_19} shows results obtained with $k=6$ and three representative settings for $p$, namely low ($p=0.1$), medium ($p=0.3$), and high ($p=0.9$) rewiring probability. In this case, WS networks are sparse, in the sense that edge density (\ref{eq:edge_density}) is low.
We note that when $p=0.1$, Fig. \ref{fig:WS11_falpha}, all three vertex observables denote right-sided asymmetry for the corresponding MFS.
As expected, multifractal characteristics are lost and in addition the spectra get shifted towards smaller $\alpha$ values as soon as the rewiring probability is increased.
In fact, for $p=0.9$ (Fig. \ref{fig:WS19_falpha}) we obtain spectra indicating monofractal character of the time series. These results have been confirmed by means of the MFDMA method \cite{PhysRevE.82.011136} (results not shown).

\begin{figure}[ht!]
\centering
\includegraphics[scale=0.7,keepaspectratio=true]{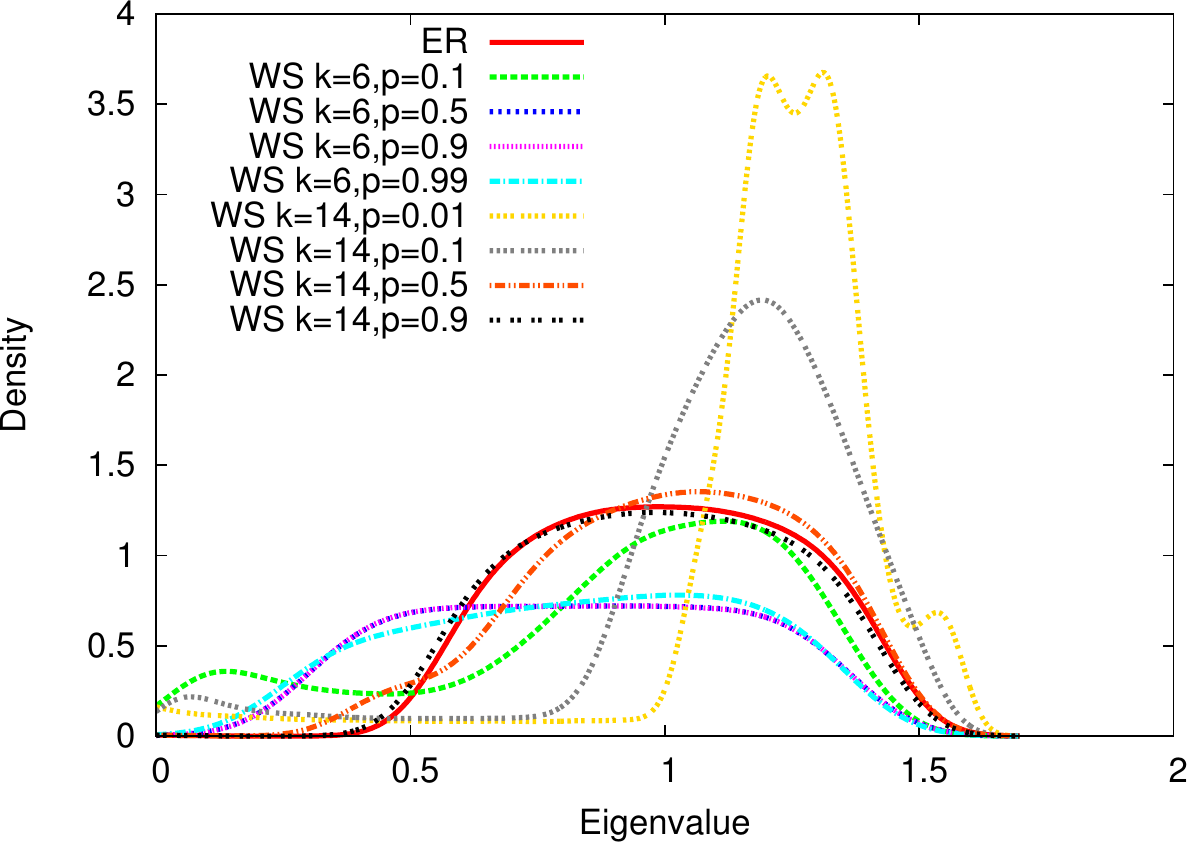}
\caption{Spectra of normalized Laplacian matrices related to WS networks generated with different model parameters. As it is possible to note, using large values for $p$ (e.g., $p=0.9$) does not yield networks whose spectra resemble the ones of ER graphs. This fact helps explaining the peculiar MFS observed for VCL time series: edge density plays an important role. In fact, the normalized Laplacian spectrum of ER graph is matched only when $k=14$ and $p\geq 0.9$.}
\label{fig:WS_NL_spectra}
\end{figure}

A peculiar behaviour is observed for VCL time series. For $p>0.3$ and $k=6$, these time series contain many zeros (data not shown) which result in spurious detection of multifractal features. Thus, even for $p=0.9$ MFS estimated for VCL time series do not resemble monofractal spectra with Hurst exponent $H\approx0.5$, as obtained for ER graphs with corresponding edge density (and equal number of vertices). In order to further elaborate on this issue, we estimated the distribution of eigenvalues of the normalized Laplacian matrices of ER graphs, as well as of different WS graphs; see Fig. \ref{fig:WS_NL_spectra}.
Our results indicate that, to obtain a normalized Laplacian spectrum of a WS network that resembles the one of a typical ER graph, such WS graphs have to be generated with large rewiring probability $p$ and also with large edge density (large $k$ considering the number of vertices).

\begin{figure}[ht!]
\centering

\subfigure[Hurst exponent vs $p$.]{
\includegraphics[scale=0.26,keepaspectratio=true]{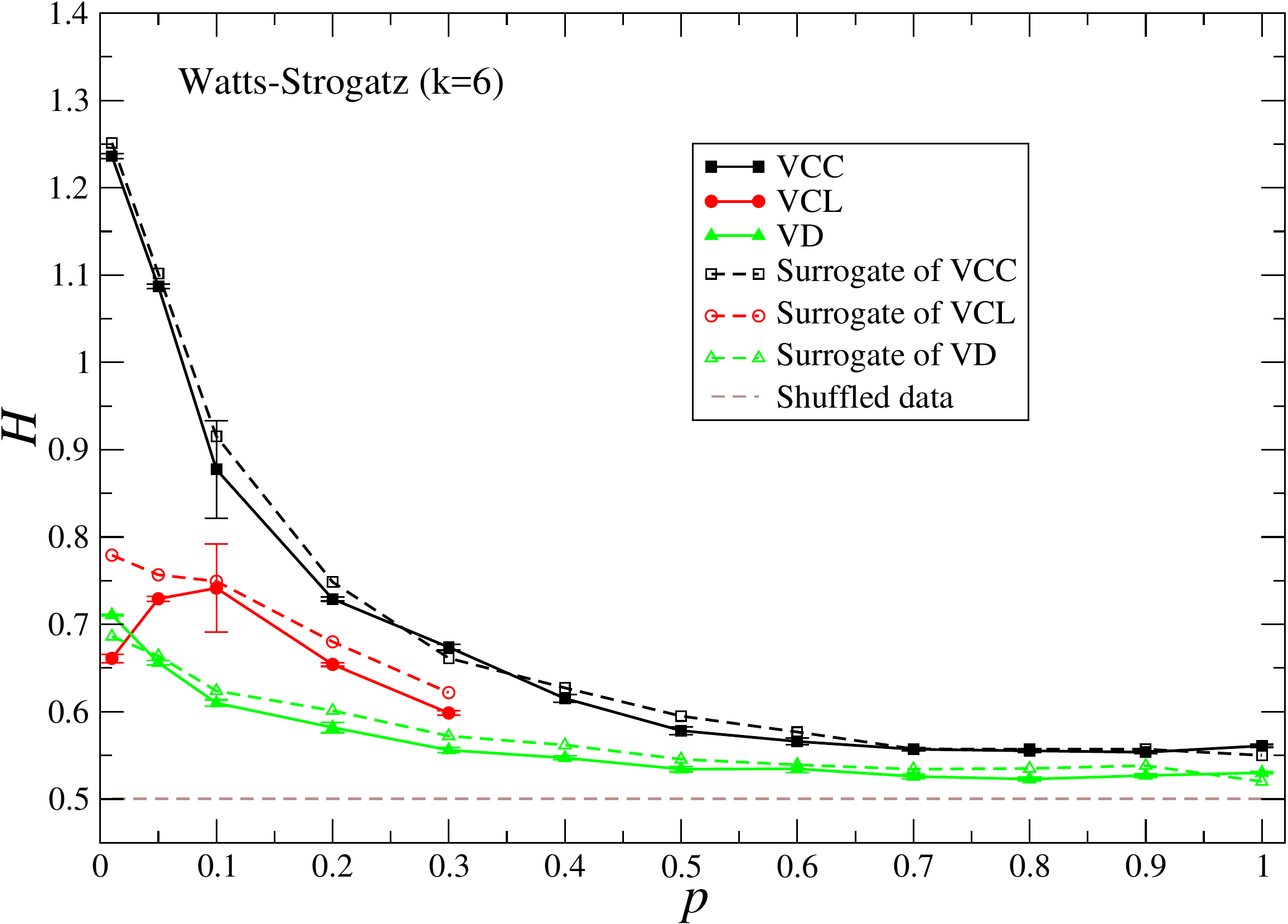}
\label{fig:WS_Hurst_k6}}
~
\subfigure[MFS width vs $p$.]{
\includegraphics[scale=0.26,keepaspectratio=true]{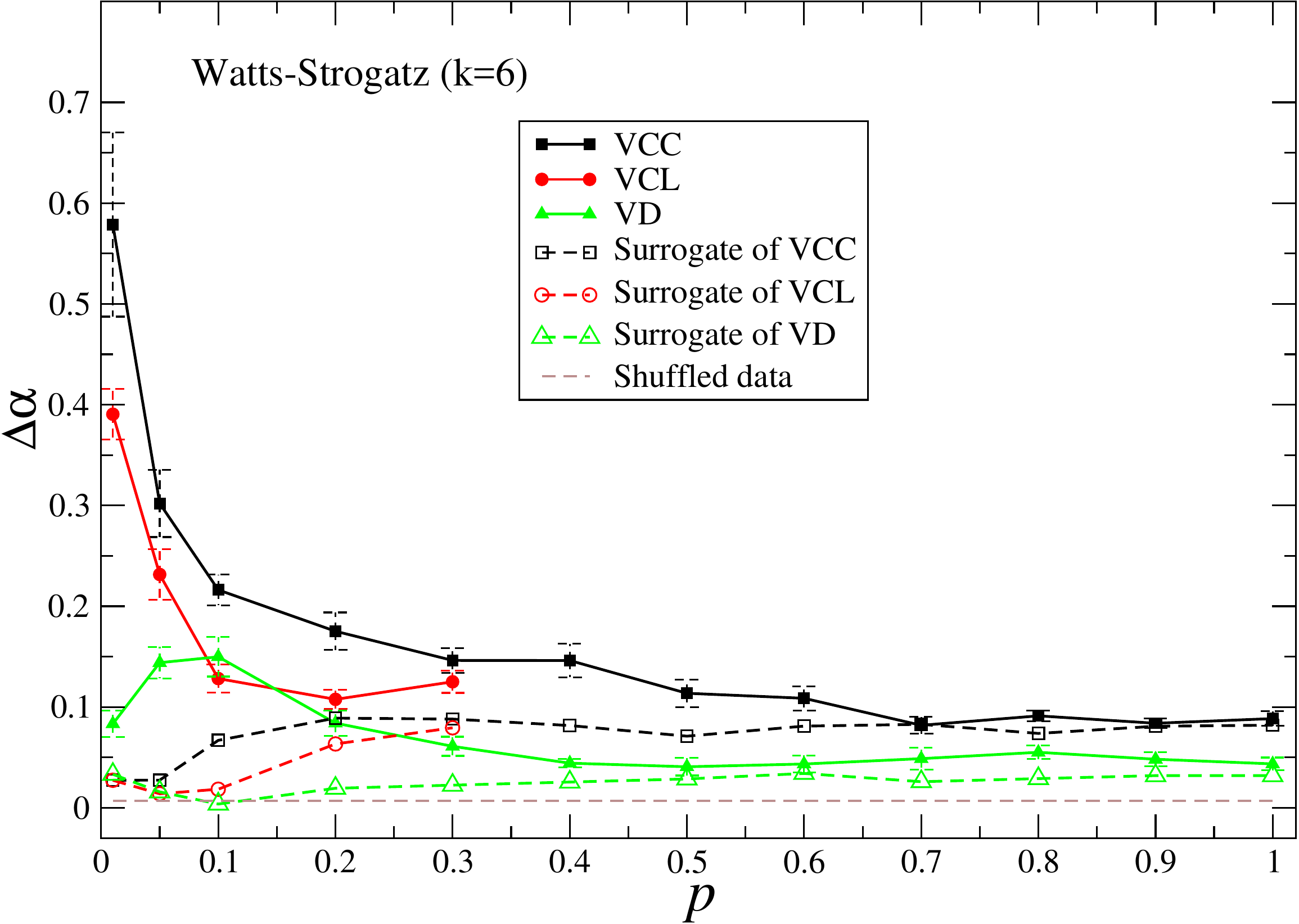}
\label{fig:WS_mfsw_k6}}

\subfigure[MFS asymmetry vs $p$.]{
\includegraphics[scale=0.28,keepaspectratio=true]{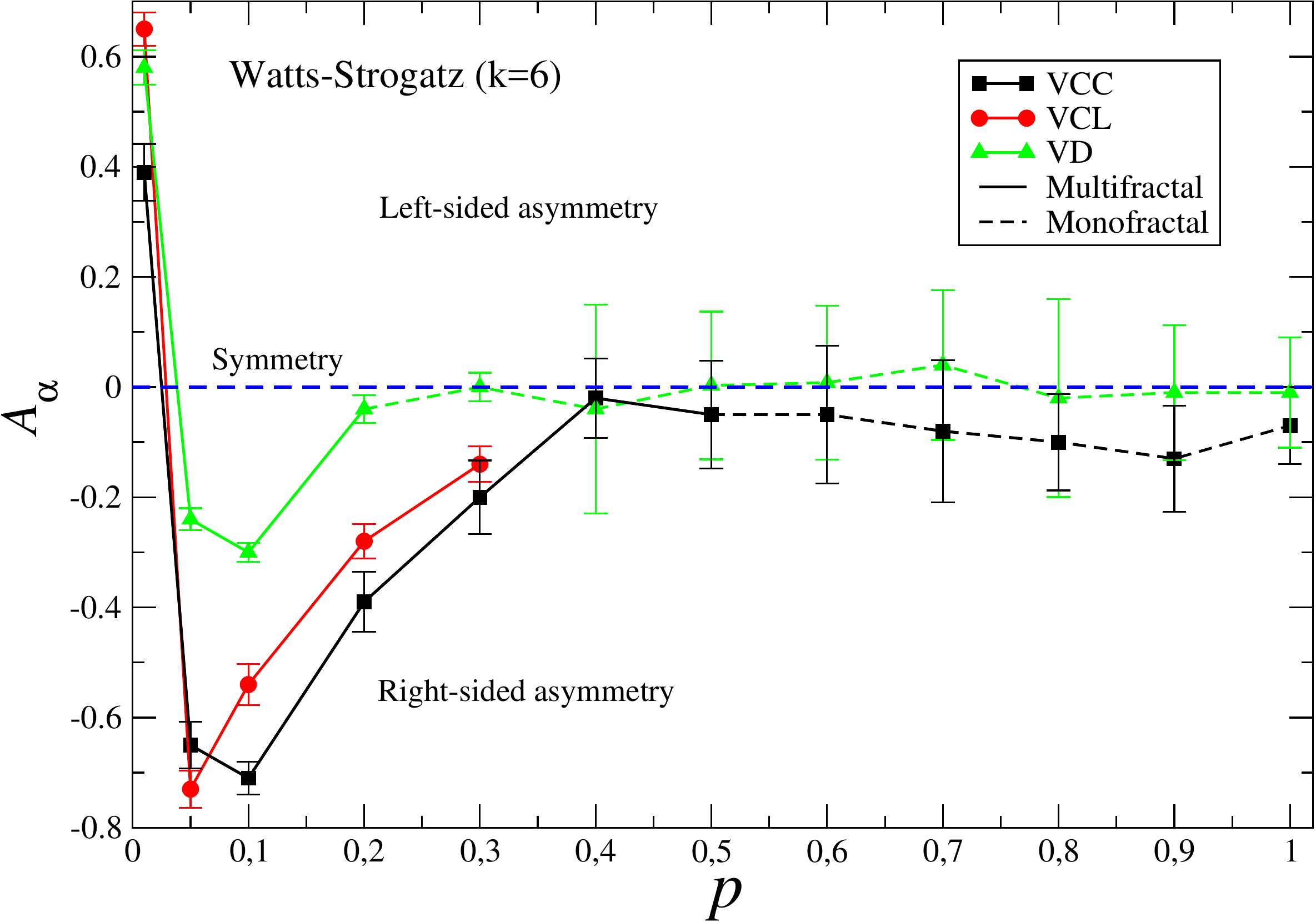}
\label{fig:WS_Asy_k6}}

\caption{Multifractal characteristics of WS graphs with $k=6$ fixed and varying $p\in[0, 1]$. Results show that, as the rewiring parameter $p$ increases, the width of MFS, as well as degree of asymmetry $A_{\alpha}$ (except for $p=0.01$) and Hurst exponent $H$, decrease systematically. Results for VCL time series and $p> 0.3$ are not shown due to the computational artefacts mentioned in the text.}
\label{fig:WS_k6}
\end{figure}

\begin{figure}[ht!]
 \centering

\subfigure[Hurst exponent vs $k$.]{
\includegraphics[scale=0.27,keepaspectratio=true]{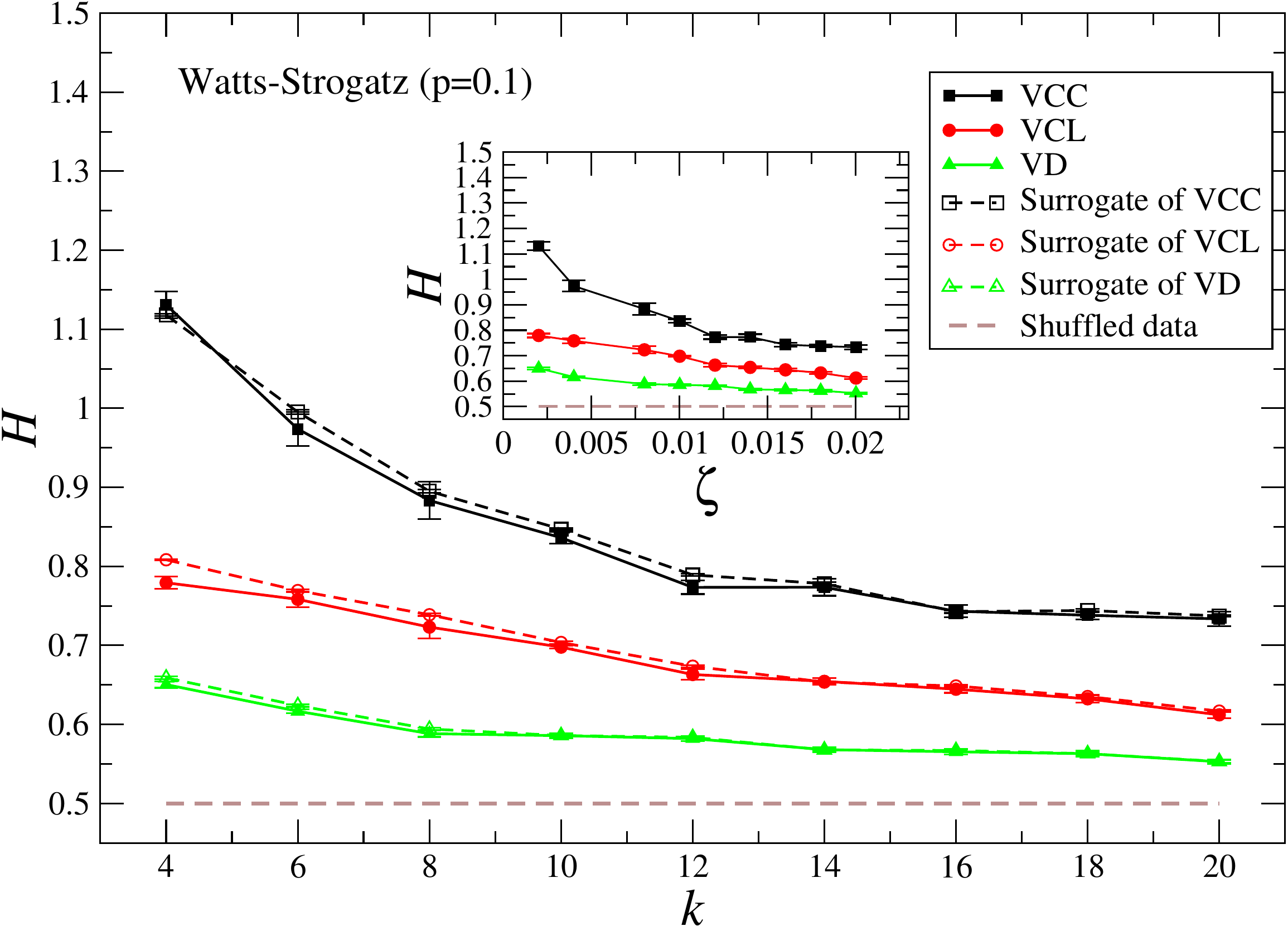}
\label{fig:WS_Hurst_p01}}
~
\subfigure[MFS width vs $k$.]{
\includegraphics[scale=0.27,keepaspectratio=true]{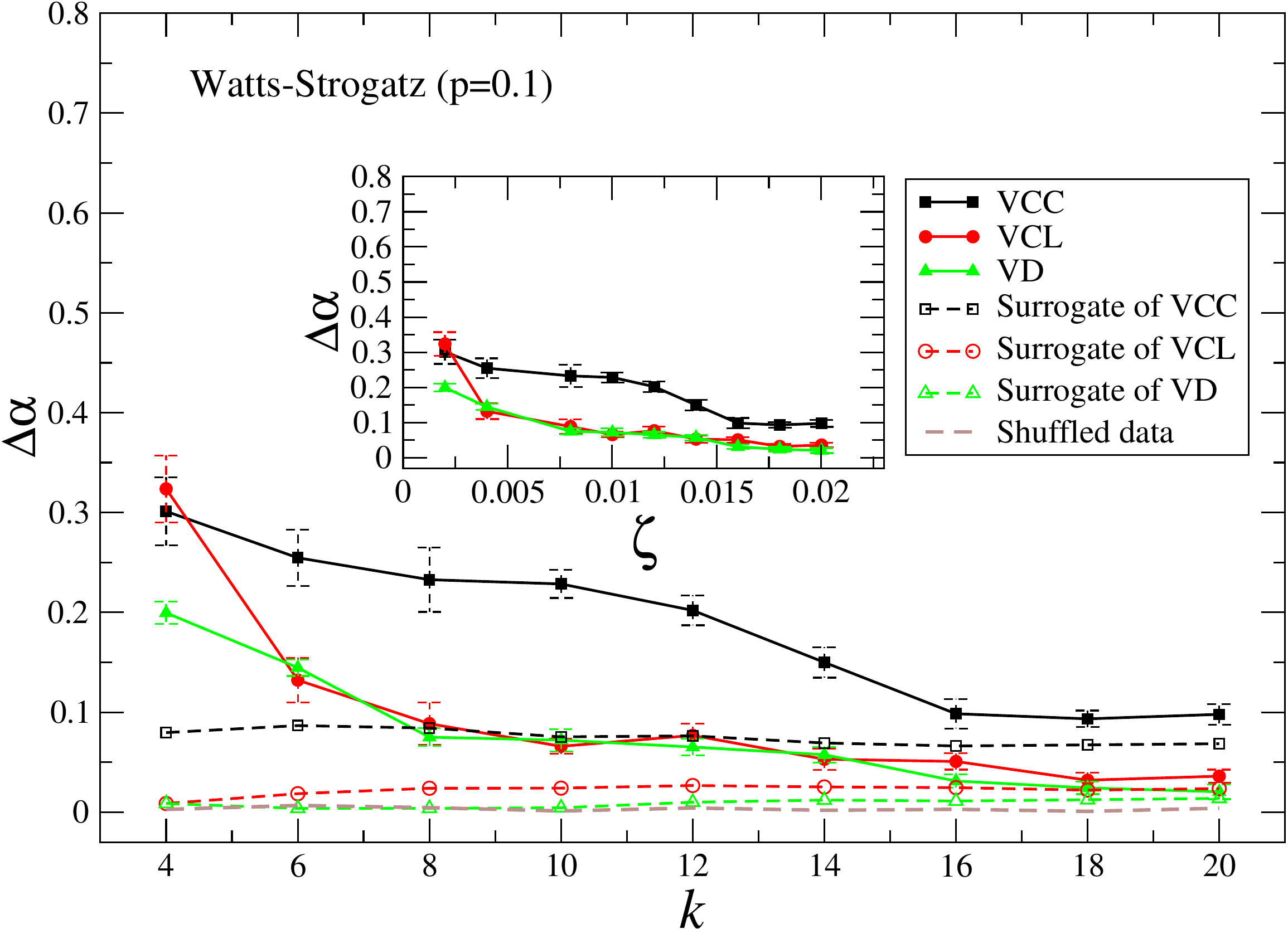}
\label{fig:WS_Dalpha_p01}}

\subfigure[MFS Asymmetry vs $k$.]{
\includegraphics[scale=0.3,keepaspectratio=true]{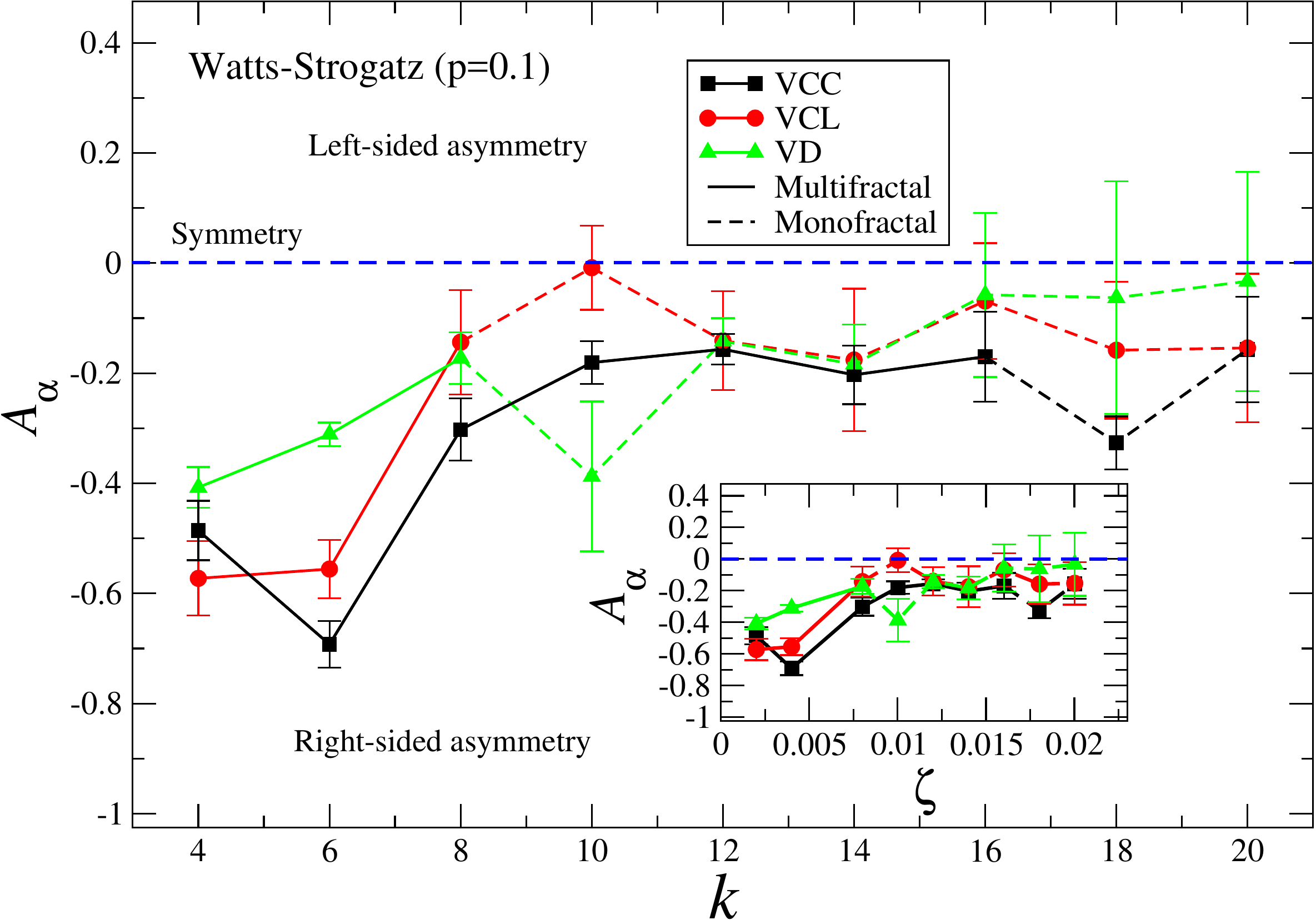}
\label{fig:WS_Asymmetry_p01}}

\caption{Multifractal characteristics of WS graphs with $p=0.1$ and varying $k\in\{4, 6, ..., 20\}$. Results show that the multifractal characteristics are dependent on the edge density. As we increase $k$, MFS width $\Delta\alpha$ decreases followed by an implicit symmetrization of the spectra. We note that, when $k\leq 8$, MFS are right-sided. These results suggest that, taken alone, small average shortest path and high clustering coefficient are not sufficient to produce a right-sided spectrum. In fact, edge density (\ref{eq:edge_density}) is a key ingredient to observe right-sided asymmetry.}
\label{fig:WS_varying_K}
\end{figure}

\begin{figure}[ht!]
 \centering
 \includegraphics[viewport=0 0 676 496,scale=0.4,keepaspectratio=true]{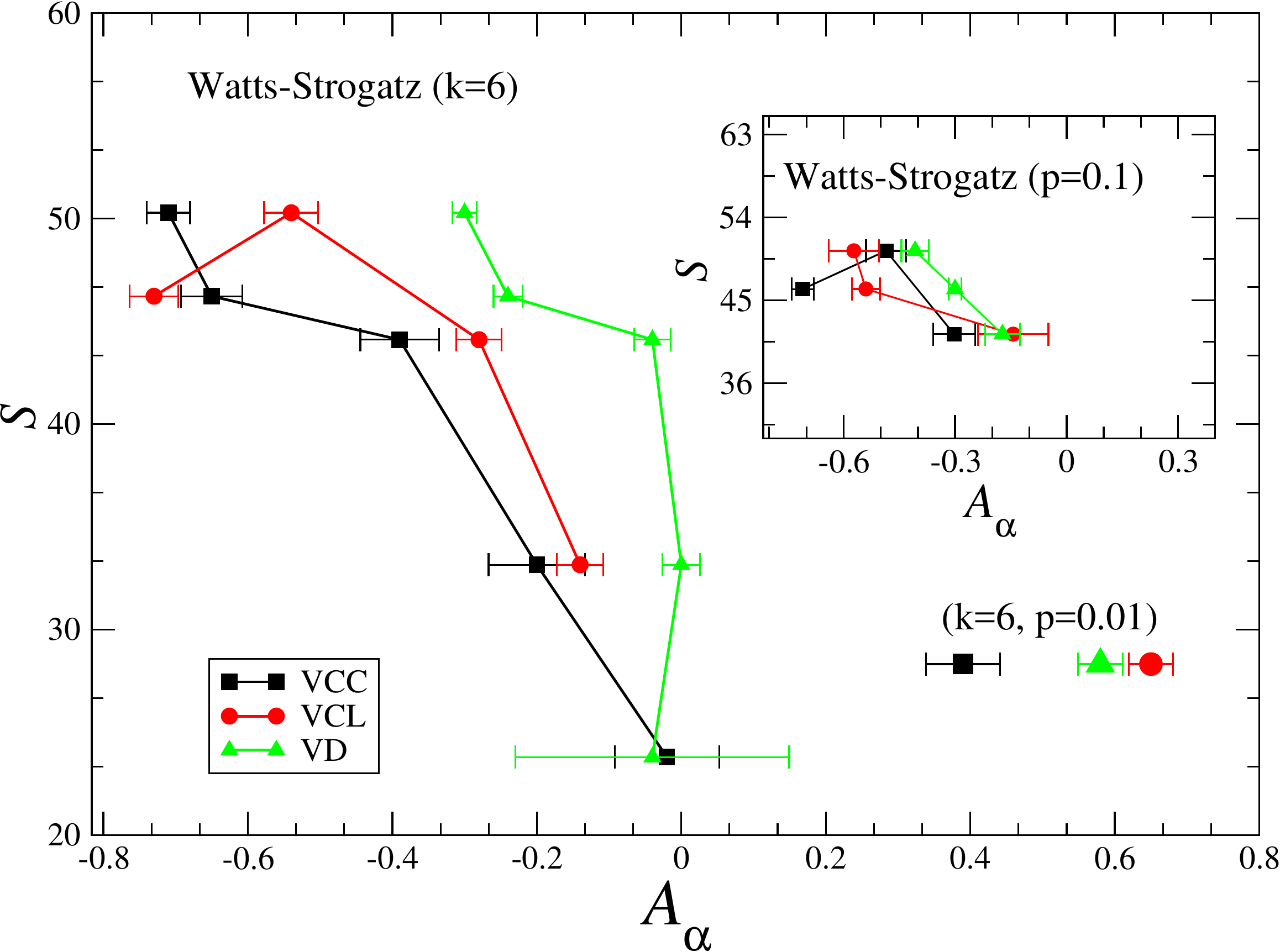}
 \caption{Plots comparing MFS asymmetry with respect to $S$ in Eq. \ref{eq:smallworldness}. According to Fig. \ref{fig:WS_k6}, this figure is showing results obtained by varying $p\in\{0.01, 0.05, 0.1, 0.2, 0.3, 0.4\}$ and fixing $k=6$. \textbf{Inset:} Results obtained according to Fig. \ref{fig:WS_Asymmetry_p01}, that is, by varying $k\in\{4, 6, 8\}$ and fixing $p=0.1$. In both cases, we report results for the range of parameter values where multifractality is clearly observed in related time series. Please notice that results for WS network obtained with $p=0.01, k=6$ are represented separately in the figure, since they give rise to wide left-sided MFS.}
 \label{fig:WS_S_Alpha}
\end{figure}

Fig. \ref{fig:WS_k6} shows how the multifractal characteristics are affected by the degree of randomness introduced in WS models by increasing $p$. In particular, we note that VD time series become monofractal when $p\geq 0.2$. On the other hand, VCC time series become monofractal when $p\geq 0.5$. 
Moreover, multifractality of the time series is mainly accompanied by a strong right-sided asymmetry of the MFS (Fig. \ref{fig:WS_Asy_k6}).
The only exceptions are spectra obtained for WS networks with a very small $p=0.01$ (the structure of such graphs is similar to the initial ring) for which we identify left-sided asymmetry of the spectrum.
The spectra shift systematically towards $H \approx 0.5$ when increasing $p$, suggesting that the temporal organization of the related time series weakens progressively when increasing degree of graph randomness. Results for VCL time series and $p>0.3$ are not shown due to the computational artefact mentioned above. Also the neighborhood size, here controlled by $k$, plays a particularly important role in determining the multifractal properties of vertex observables; see Fig. \ref{fig:WS_varying_K}. Our results show that edge density is a key ingredient for obtaining MFS with right-sided asymmetry. In fact, regardless of the values assigned to $p$, when edge density is too high, MFS become narrow, i.e., monofractal.
Fig. \ref{fig:WS_S_Alpha} shows the relationship between MFS asymmetry and the small-worldness measure discussed in Sec. \ref{sec:smallworldness}.
We note that for $p=0.05, k=6$, the asymmetry degree becomes negative, indicating an asymmetric spectrum stretched on the right-hand side. The value for $S$ in Eq. \ref{eq:smallworldness} are in agreement with this behavior. In fact, in the $p=0.01$ case, we have $S=28.316$, while in the $p=0.05$ case, we have a higher value, i.e., $S=46.219$.

These results, taken altogether, suggest that right-sided asymmetry of MFS could be interpreted also as a signature of network small-worldness.
Notably, WS models producing time series with right-sided spectra are more small-world than WS models denoting either monofractal or left-sided MFS asymmetry in the corresponding time series.
Finally, it is worth mentioning that differences between $S$ values calculated for different network configurations (hence yielding different values for $A_{\alpha}$) are all statistically significant (details not shown).

\begin{figure}[ht!]
\centering

\subfigure[Time series.]{
\includegraphics[scale=0.3,keepaspectratio=true]{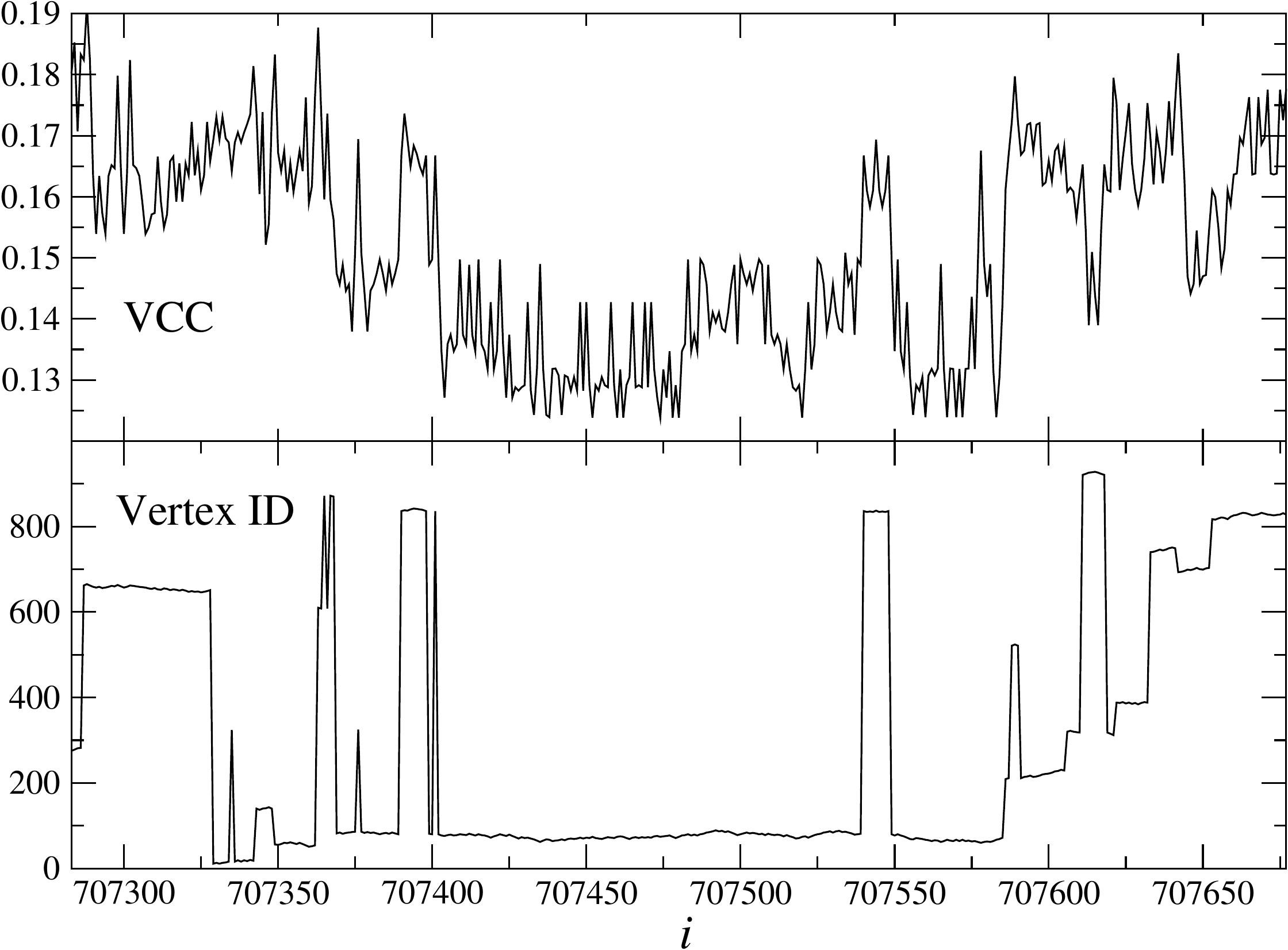}
\label{fig:WS11_sample_ts}}

\subfigure[Graph visualization.]{
\includegraphics[scale=0.8,keepaspectratio=true]{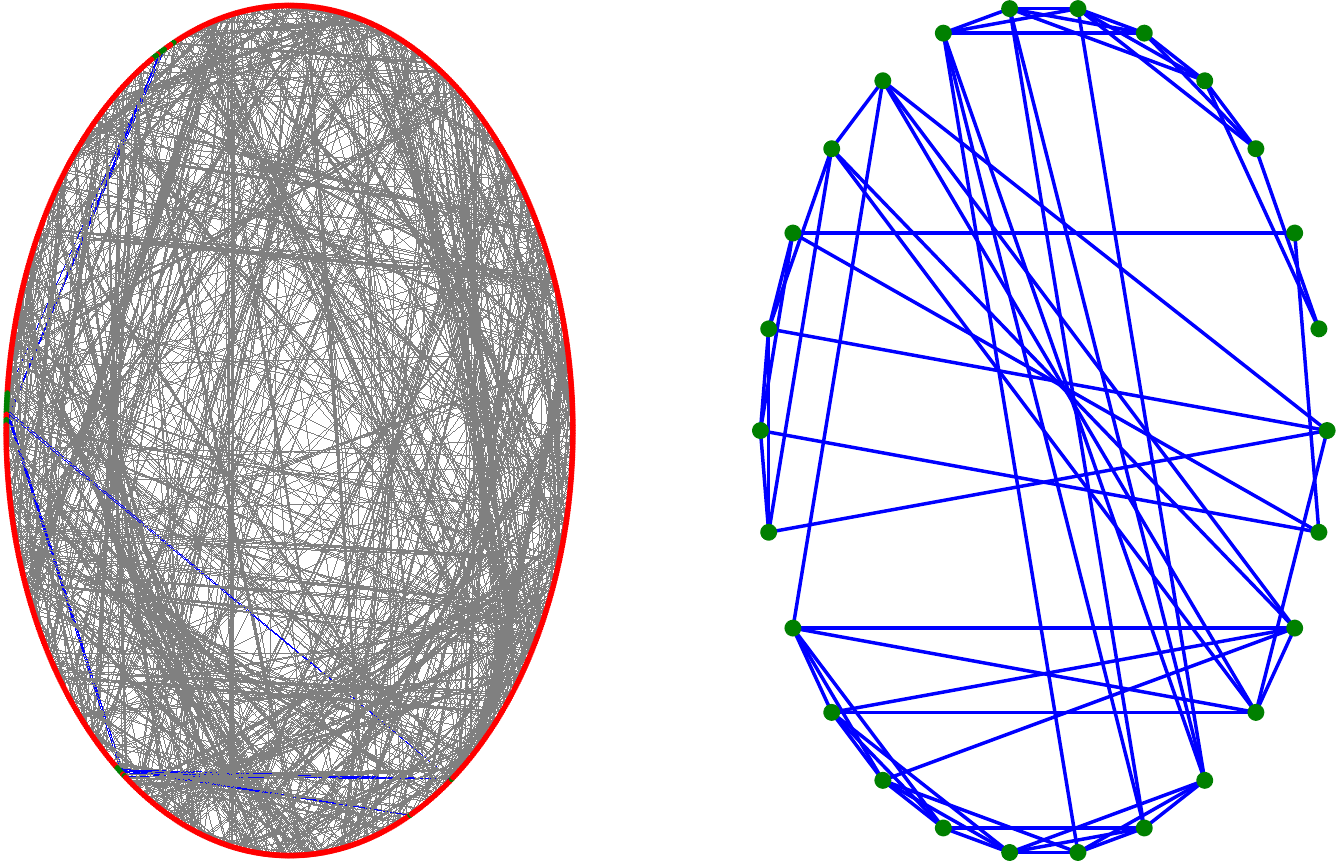}
\label{fig:WS11_graph}}
 
\caption{Results obtained for WS model with $p=0.1$. A fragment of the VCC time series with corresponding vertex identifiers is shown in Fig. \ref{fig:WS11_sample_ts}. In Fig. \ref{fig:WS11_graph}, we show a circular visualization of the entire WS network and a magnification corresponding to the time series fragment taken into account. Edges in blue denote links visited during the random walk.}
\label{fig:WS11_sample}
\end{figure}

The qualitative explanation of the MFS asymmetry can be inferred from Fig. \ref{fig:WS11_sample}, where we show a sample time series (Fig. \ref{fig:WS11_sample_ts}) obtained for WS network with $p=0.1$ as well as a visualization of the network (Fig. \ref{fig:WS11_graph}). It can be easily noticed that large fluctuations of the closeness centrality coefficient, VCC, are accompanied with long-range network links. Moreover, long-range links are randomly distributed and constitute a small fraction of all connections in the graph. Hence, the random walker visits mainly short-range connected vertices for which centrality coefficient fluctuates in a complex way, yet without abrupt changes. This, in turn, results in VCC time series having a more complex temporal organization for small fluctuations rather than for the large ones, which is manifested by right-sided asymmetry of the related MF spectrum.

Robustness of our analysis is demonstrated by reporting results for surrogate time series \cite{schreiber2000surrogate}.
For this purpose, we considered two kinds of tests commonly used in the statistical analysis of time series. The former relies on analysis of artificial time series generated by means of phase-randomized Fourier transform algorithm applied to the original data. Hence, by means of this procedure one can destroy potential non-linear autocorrelations existing in the time series but still preserve the linear ones.
The latter test is based on shuffling the original time series. This procedure is a simple and efficient way for destroying all temporal correlations yet preserving the probability density function of the data. Results on surrogates confirm statistical significance of our analysis. In the case of surrogates based on Fourier transform, only monofractal behaviour ($\Delta \alpha < 0.1$) \cite{drozdz2016} is identified, whereas linear dependencies related to Hurst exponent remain unchanged. Similarly, randomly shuffled data are characterized by homogeneous scaling properties being quantified by a single exponent $h(q)= H= 0.5$ and extremely narrow spectrum MFS ($\Delta \alpha \approx 0$).

\begin{figure}[ht!]
 \centering
 \includegraphics[scale=0.4,keepaspectratio=true]{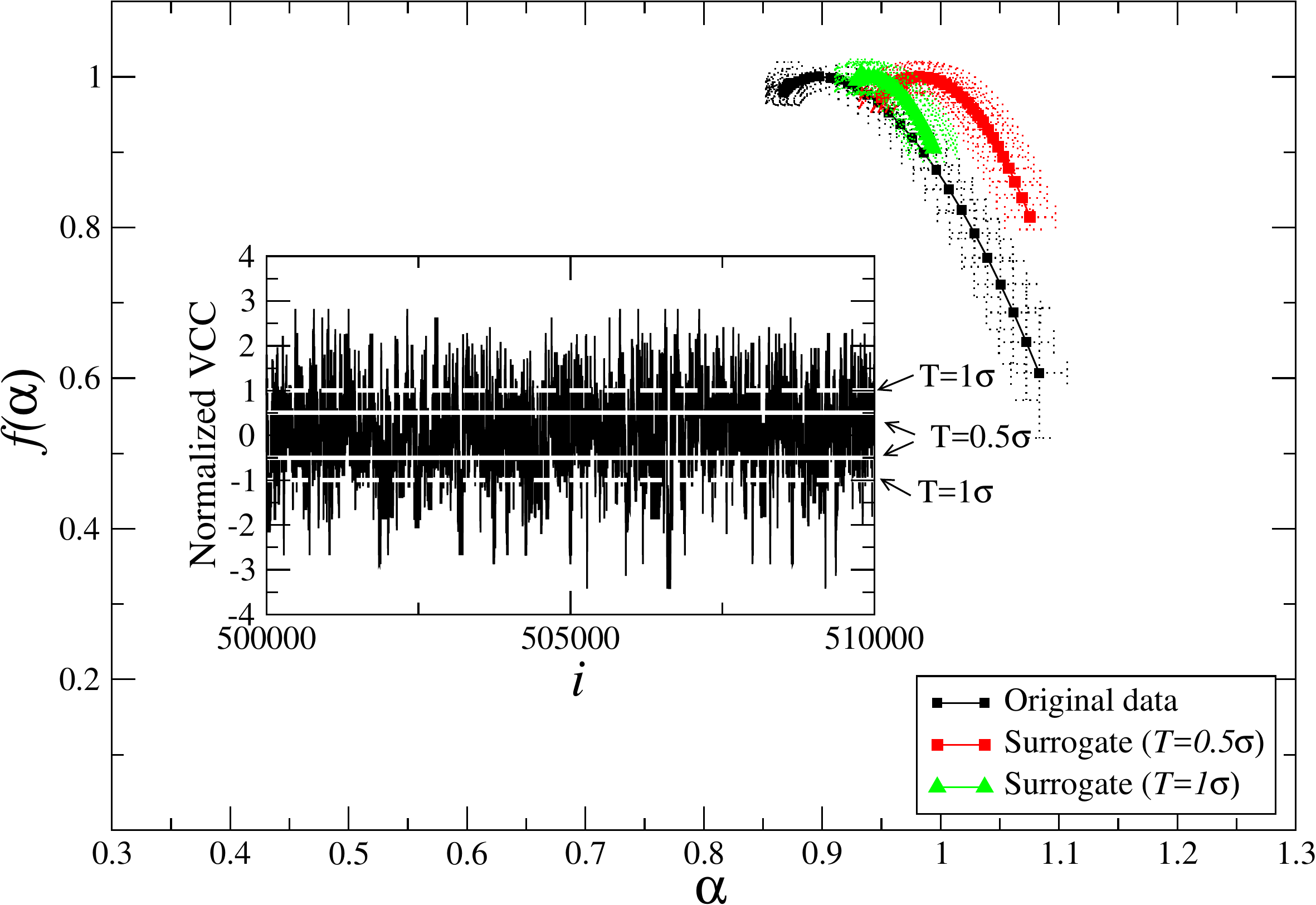}
 \caption{\textbf{Main:} MFS spectra for surrogate of VCC time series for two values of threshold $T=\{0.5\sigma,1\sigma \}$. Monofractal spectrum is obtained for $T=1\sigma$, confirming that multifractality results from the temporal organization of the small and medium fluctuations. \textbf{Inset:} Sample fragment of the VCC time series with threshold level marked with white lines.}
 \label{fig:WS_A_surrogate}
\end{figure}

We also perform additional tests to illustrate the contribution of the temporal organization of the small and medium fluctuations in the right-sided asymmetry of MFS. To this end, data points of original time series with amplitude smaller than a threshold are randomized, whereas larger ones remain unchanged. This procedure is equivalent to removing correlations between small fluctuations, leaving the larger ones intact.
We apply this procedure to the VCC time series related to the WS model with $p=0.1$, which are characterized by strong MFS right-sided asymmetry. For normalized time series (zero mean and unit variance), we consider two values for the threshold $T=\{0.5\sigma, 1\sigma \}$. The results are depicted in Figure \ref{fig:WS_A_surrogate}.
It is evident that random shuffling of the small and medium fluctuations destroys multifractality in the time series. For $T=1\sigma$ we obtain monofractal spectrum ($\Delta \alpha \approx 0.07$), confirming that complexity of the VCC time series is related to temporal organization of the small fluctuations.


%%%%%%%%%%%%%%%%%%%%%%%%%%%%%%%%%%%%%%
\clearpage
\subsection{Dorogovtsev-Goltsev-Mendes model}

\begin{figure}[ht!]
\centering

\subfigure[Fluctuation functions for DGM9.]{
\includegraphics[scale=0.32,keepaspectratio=true]{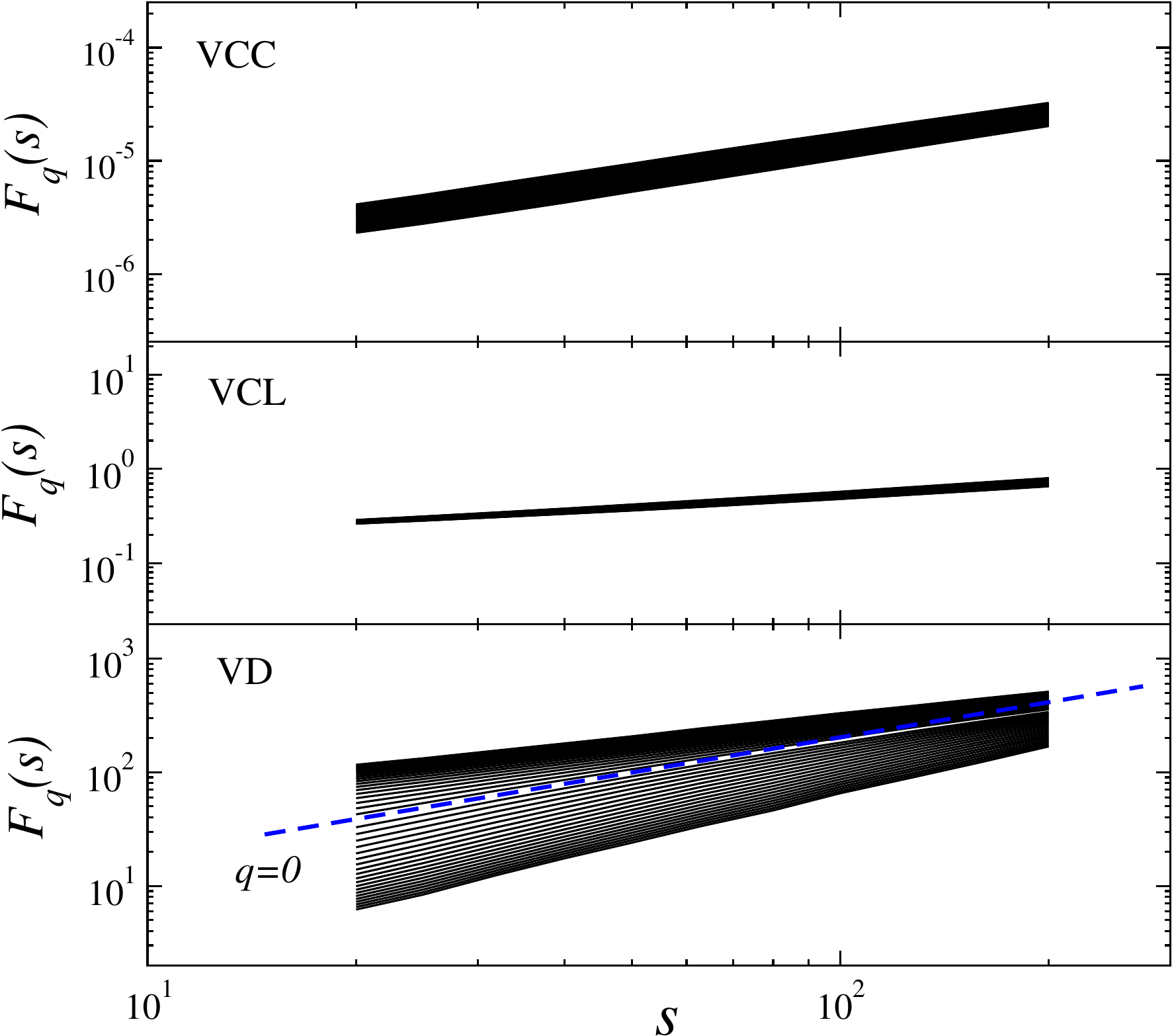}
\label{fig:DGM9_Fq}}
~
\subfigure[MFS for DGM7, DGM8 and DGM9.]{
\includegraphics[scale=0.29,keepaspectratio=true]{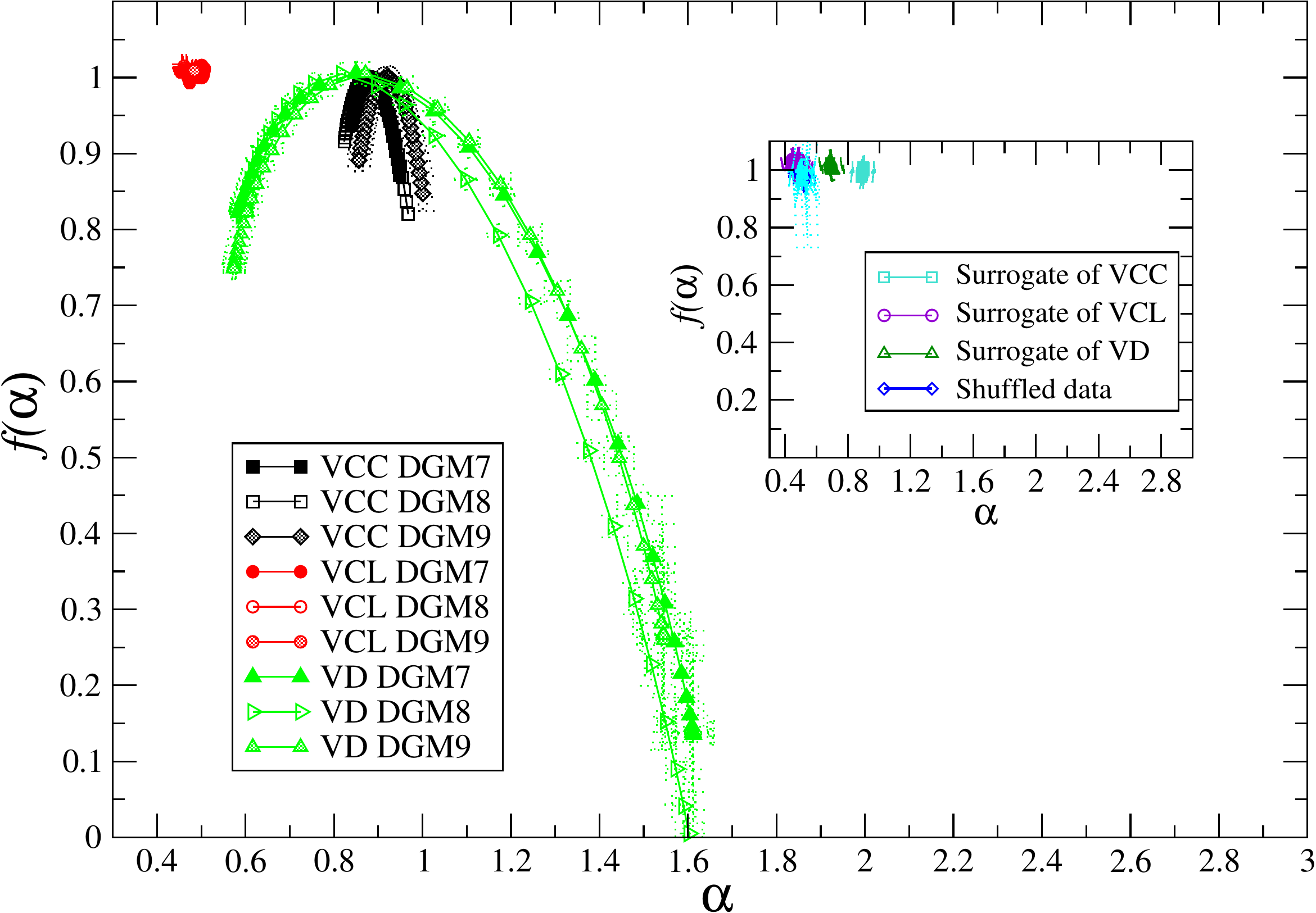}
\label{fig:DGM_falfa}}

\caption{Results for DGM7, DGM8, and DGM9. Only VD time series denote multifractal characteristics. Related MFS are right-sided with a similar degree of asymmetry.}
\label{fig:DGM}
\end{figure}

In this section, we analyze results obtained by means of a scale-free network model introduced by Dorogovtsev, Goltsev, Mendes (DGM) \cite{dorogovtsev2002pseudofractal}.
The DGM model is deterministic and generates scale-free networks whose average shortest path grows logarithmically with the number of vertices. Hence, DGM models are also small-world.
We consider three DGM model instances that differ by the number of iterations (7, 8, and 9) used during the growth process: resulting networks are denoted with DGM7, DGM8, and DGM9, respectively.
Results are shown in Fig. \ref{fig:DGM}. It is possible to note that multifractality is observed only for VD time series, where MFS are wide and strongly right-sided.
The asymmetry is the same for all DGM networks taken into account, $A_{\alpha}\simeq -0,473$. However, $S$ in Eq. \ref{eq:smallworldness} changes and increases significantly: for DGM7 is 244.41; for DGM8 is 1153.88; and for DGM9 is 2709.81. This fact indicates that the signature of small-worldness (\ref{eq:smallworldness}) should increase in DGM networks with the size of the graphs. However, this should not happen as the average shortest path scaling law remains the same regardless of the graph size. This might suggest that $S$ is sensitive to the network size and related edge density (DGM7 contains 1095 vertices and 2187 edges, which gives $\zeta=0.003$; $\zeta=0.001$ for DGM8; $\zeta=0.0004$ for DGM9).

Nonlinear dependence between data constitute the main factors responsible for multifractal organization of time series \cite{drozdz2009quantitative}.
However, the fat-tailed distribution of data can also be a significant ingredient of multifractality when nonlinear correlations are present \cite{zhou2009components}.
A transformation applied to a given time series preserves the hierarchical organization of data but changes the underlying distribution. This, in turn, might result in changes of the correlation structure of data that, in turn, depend on the change of the distribution.
In the DGM case, VD and VCL are linked by a precise relation: $c_i=2/k_i$.
The distribution of the VD time series (variance $\sim 3700$) is much more dispersed than VCL ones (variance $\sim 0.15$).
Thus, heterogeneity of correlation organization between fluctuation of different amplitudes reflected in multifractal characteristics is much more evident for VD data than for VCL time series.

\subsection{Song-Havlin-Makse fractal model}
\label{sec:shm}

The Song-Havlin-Makse (SHM) model \cite{song2006origins} allows to generate networks with modular topologies ranging from fractal to non-fractal networks having small ASP.
The SHM model comprises of four main parameters: (i) number of generations, (ii) number of offspring per vertex, (iii) number of connections between offsprings, (iv) and $e$, the probability that hubs stay connected during the growth process. Here we keep fixed the first three parameters with values 4, 3, and 2, respectively, and vary $e$ within the $[0, 1]$ range in order to study the effects on the related time series of vertex observables in terms of MFS features.
When $e<1$, hubs grow by preferentially linking with low-degree vertices (anticorrelated attachment mechanism), giving rise to a robust fractal topology.
Attraction between hubs obtained with large values of $e$, instead, leads to non-fractal topologies with low ASP -- a key ingredient in small-world networks.
In all SHM models, VCL is always zero.

\begin{figure}[ht!]
\centering

\subfigure[Hurst exponent vs $e$.]{
\includegraphics[scale=0.27,keepaspectratio=true]{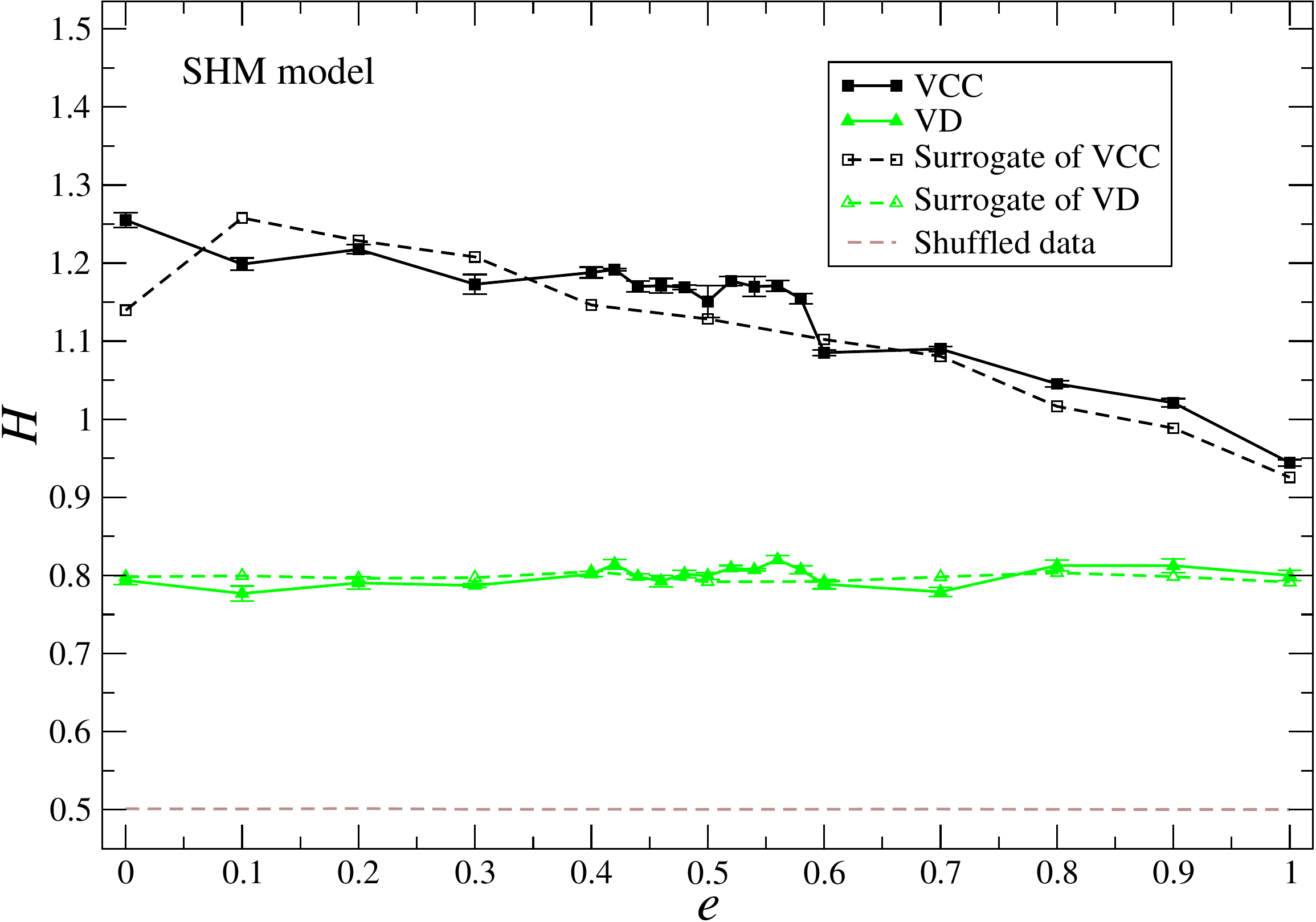}
\label{fig:SHM_Hurst}}
~
\subfigure[MFS width vs $e$.]{
\includegraphics[scale=0.27,keepaspectratio=true]{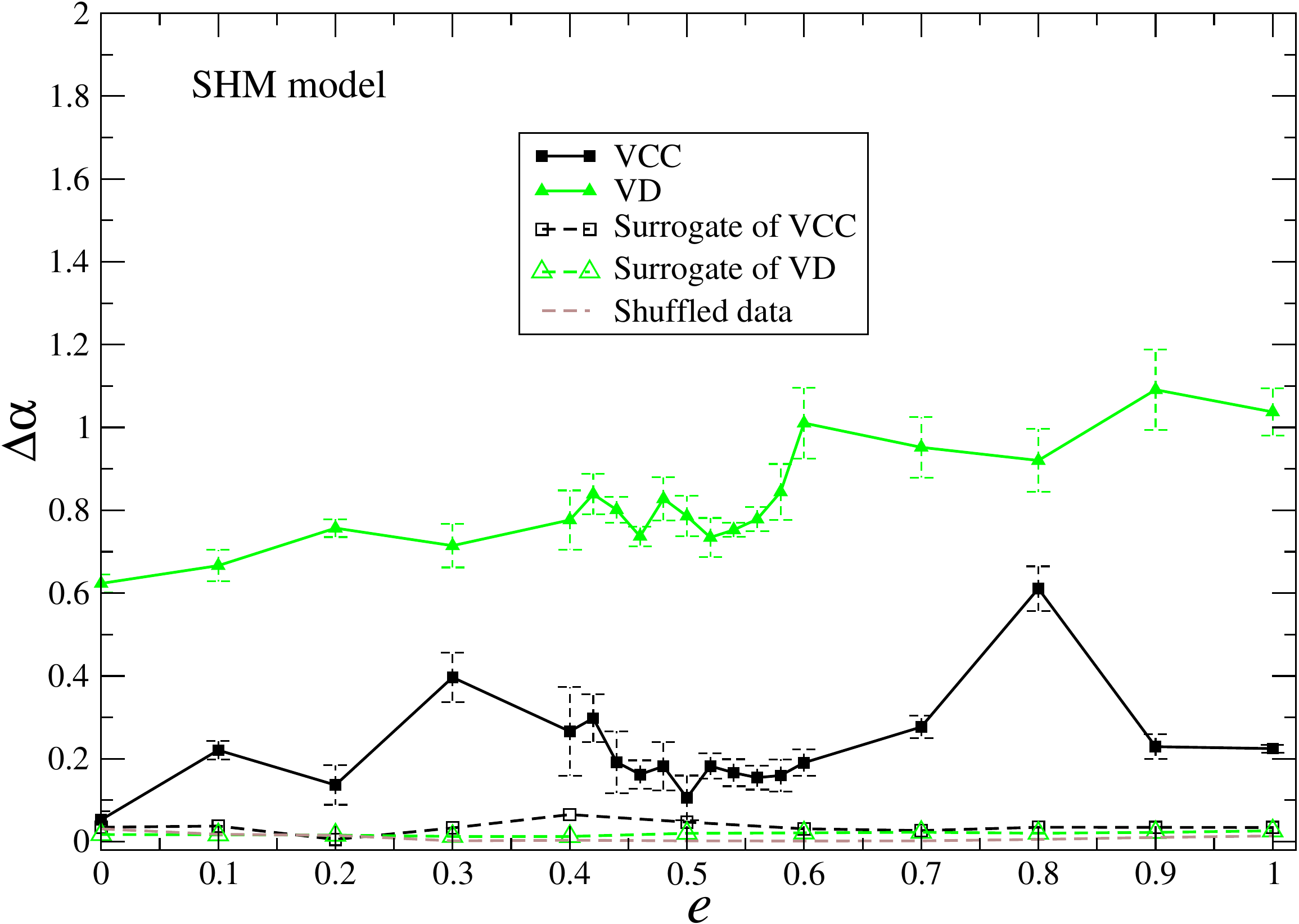}
\label{fig:SHM_Dalpha}}

\subfigure[Asymmetry vs $e$.]{
\includegraphics[scale=0.26,keepaspectratio=true]{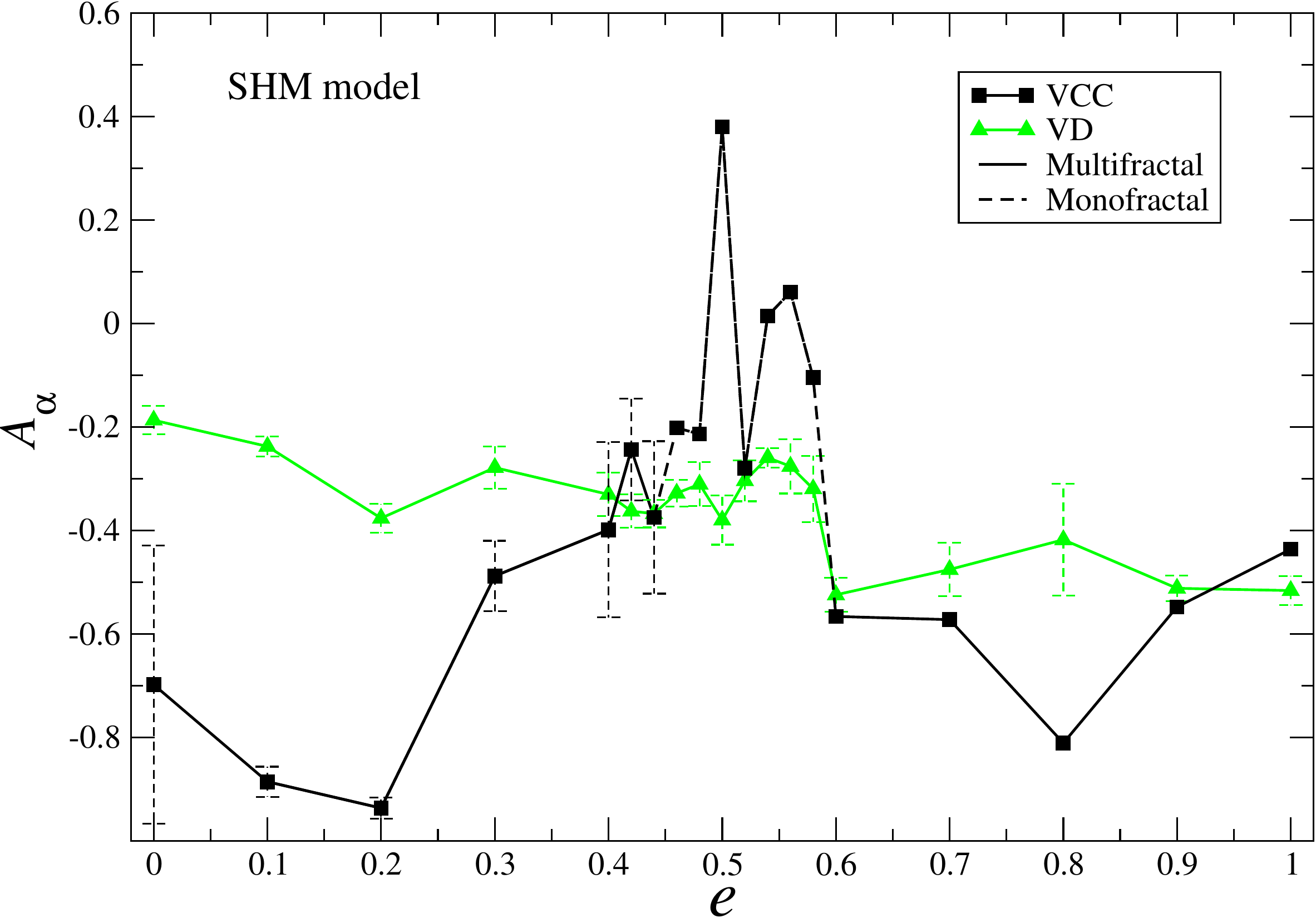}
\label{fig:SHM_Asymmetry}}
~
\subfigure[Position of the cross-over for VCC time series ($s_x$ indicates the scales).]{
\includegraphics[scale=0.26,keepaspectratio=true]{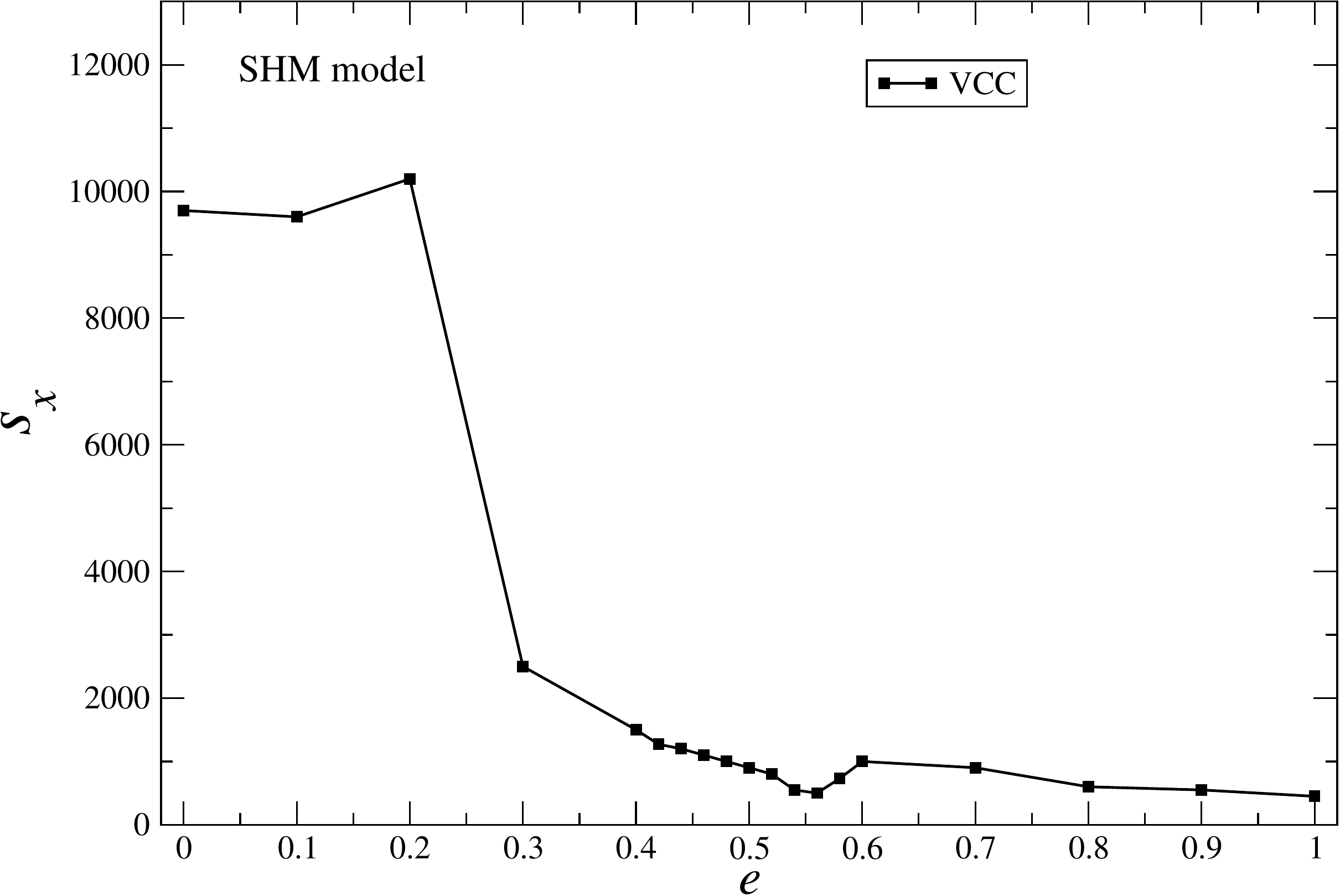}
\label{fig:SHM_crossover}}

\caption{Multifractal characteristics for SHM model with parameter $e$ (probability that hubs stay connected) varying in the $[0, 1]$ range. (a) VD time series denote nearly constant Hurst exponent as $e$ changes, whereas $H$ for VCC time series decreases when increasing $e$. (b) For VD time series, MFS width increases for network with stronger signature of small-worldness (larger $e$). For VCC case, the multifractality is observed only for $e\approx 0.3$ and $0.8$. (c) For VD time series, right-sided asymmetry of MFS increases with the degree of small-worldness. Asymmetry of the spectrum for VCC time series denotes stronger irregularity. (d) Position of the cross-over in the fluctuation functions for VCC case is dependent on the $e$ values, indicating its relation with the degree of self-similarity of the network.}
\label{fig:SHM_varying_e}
\end{figure}

In Fig. \ref{fig:SHM_Hurst}, we show the results for Hurst exponent. The VD time series denote a nearly constant Hurst exponent as we vary $e$.
The Hurst exponents of VCC time series are characterized by a decreasing trend, remaining strongly persistent for the entire parameter range.
Fig. \ref{fig:SHM_Dalpha} shows the MFS width for the two vertex observables. As it is possible to notice, for VD time series the width increases as we generate networks possessing a stronger small-world signature (i.e., by increasing $e$). Accordingly, it is possible to claim that VD time series exhibit multifractal characteristics by varying $e$ in the entire $[0, 1]$ range.
In the VCC case, instead, we note a much stronger irregularity in terms of multifractal characteristics.
A clear multifractal signature is observed when $e$ is roughly 0.3 and 0.8, whereas in the other cases MFS width suggest monofractal behavior.

Fig. \ref{fig:SHM_Asymmetry} shows the results for MFS asymmetry. For VD time series, we observe an increase in the degree of right-sided asymmetry as the networks become more small-world. Therefore, in this case the degree of right-sided asymmetry is a good predictor of the related network small-worldness.
On the other hand, for the VCC time series we observe a significant change of asymmetry when $0.45 \leq e \leq 0.58$. Notably, MFS become strongly left-sided.
However, the related MFS are narrow in that parameter range, denoting monofractal characteristics. Hence, we conclude that asymmetry of VCC time series does not seem to be useful to predict the small-worldness of the related SHM networks.

Finally, it is worth noting that, as shown in Fig. \ref{fig:SHM_crossover}, the position of the cross-over for VCC time series changes as we vary the model parameter $e$. In particular, when SHM networks are mostly fractals, i.e., when $e\simeq 0$, it is possible to analyze also larger scales (i.e., $s>10^3$) in the related time series. In fact, in all other network models taken into account in our study, we always detected a cross-over roughly at $s\simeq 10^3$, leading to uncorrelated behavior.
We hypothesize that this fact is linked with the self-similarity of the SHM network topology and leave further investigations as future work.

\begin{figure}[ht!]
\centering

\subfigure[Pure fractal.]{
\includegraphics[scale=0.27,keepaspectratio=true]{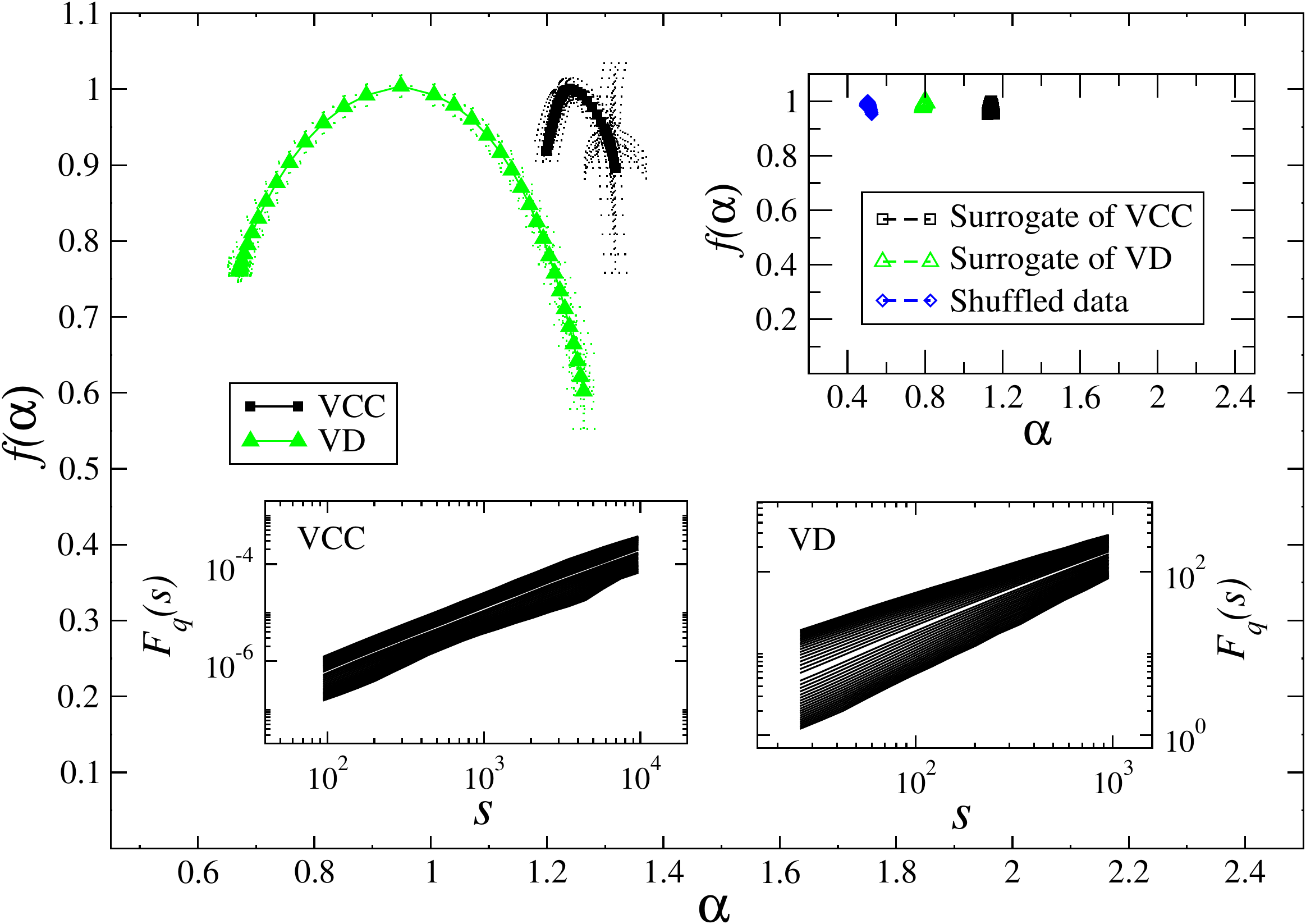}
\label{fig:SHM0_falfa_Fq}}
~
\subfigure[Hybrid network.]{
\includegraphics[scale=0.27,keepaspectratio=true]{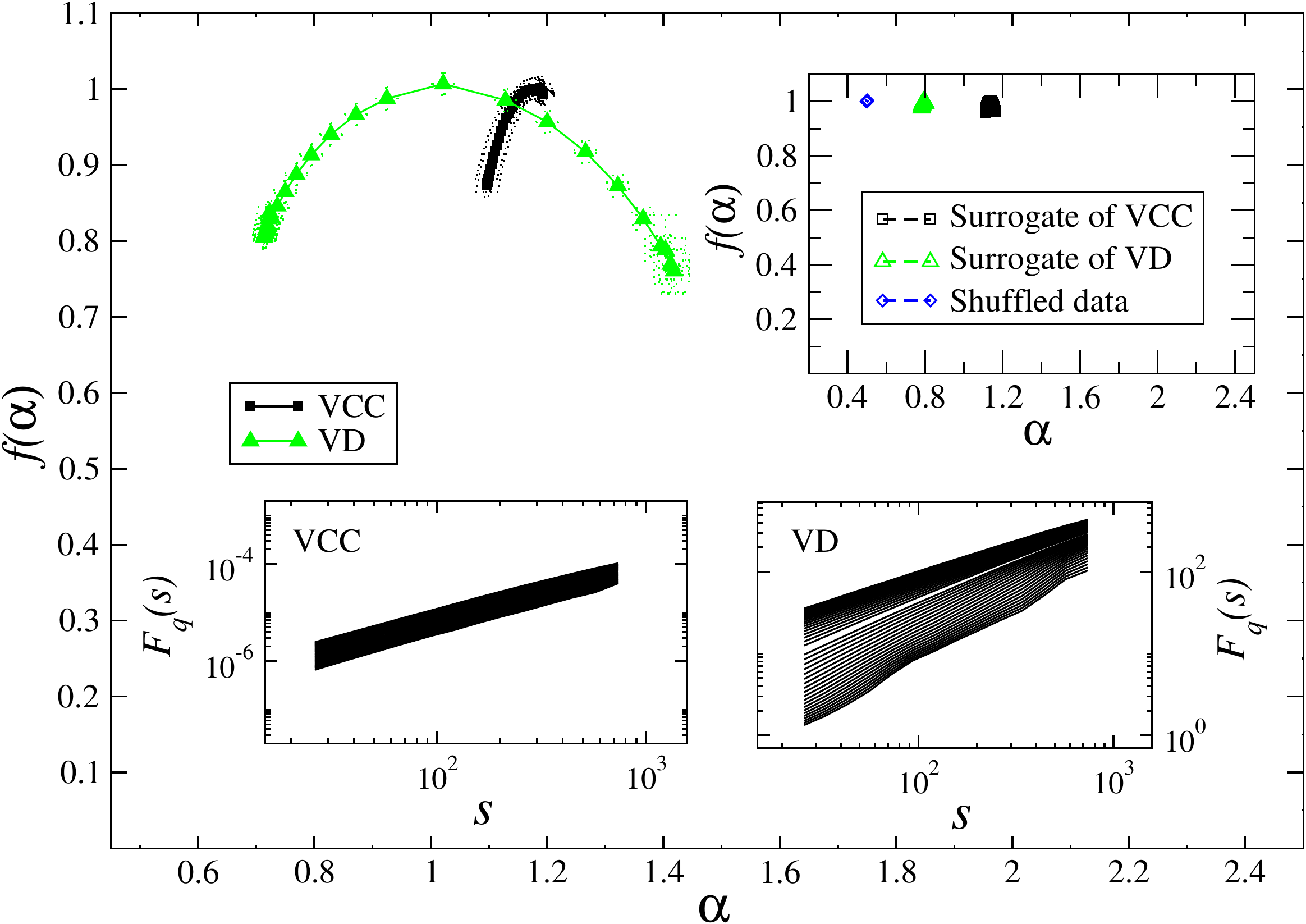}
\label{fig:SHM05_falfa_Fq}}

\subfigure[Pure small-world.]{
\includegraphics[scale=0.3,keepaspectratio=true]{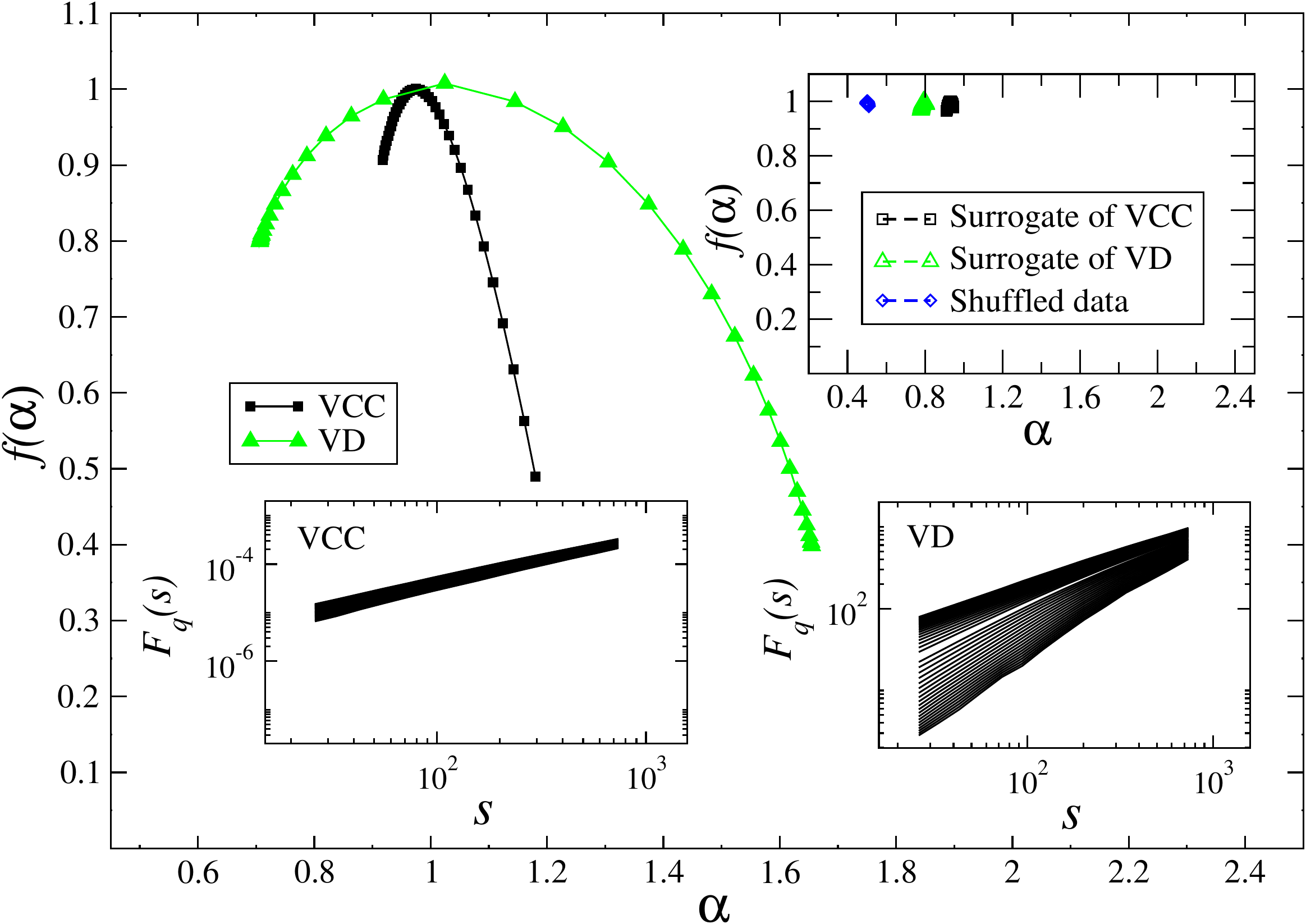}
\label{fig:SHM1_falfa_Fq}}

\caption{Results for SHM model by considering $e\in\{1, 0.5, 0\}$. Right-sided asymmetry of the MFS is visible only for pure small-world networks ($e=0$).}
\label{fig:SHM}
\end{figure}

In Fig. \ref{fig:SHM}, we show the details for three network configurations giving rise to pure fractal (a), hybrid (b), and small-world (c) topologies.
Let us first discuss the results for VCC time series.
For pure fractal models, shown Fig. \ref{fig:SHM0_falfa_Fq}, time series are homogeneous fractal with narrow and symmetric MFS ($\Delta \alpha \simeq 0.1$). Moreover, VCC time series are strongly persistent (large Hurst exponent), which we hypothesize to be related to the ``forest'' structure of networks giving rise to strong assortativity \cite{newman2002assortative} of closeness centrality ($\simeq 0.9)$.
In the case of hybrid networks, shown in Fig. \ref{fig:SHM05_falfa_Fq}, we note left-sided asymmetry of MFS (although $\Delta \alpha \simeq 0.1$). In this case, Hurst exponent is $1.16$ and assortativity of VCC is high (0.88).
Finally, Fig. \ref{fig:SHM1_falfa_Fq} shows results for pure small-world networks, where we observe strong right-sided asymmetry of multifractal spectrum with large Hurst exponent $\simeq 0.95$.

The MFS denote different characteristics for the three configurations also in the case of VD time series.
In general, $f(\alpha)$ becomes systematically wider (and more right-sided asymmetrical) as the networks become more small-world (i.e., by increasing $e$).
However, it is worth noting that the MFS shown in Fig. \ref{fig:SHM1_falfa_Fq} are different from those shown in Fig. \ref{fig:WS11_falpha}.
In fact, SHM models with large $e$ posses wider MFS with a more pronounced right-sided asymmetry than WS networks.

Overall, these results strengthen the hypothesis that small-world topologies with low edge density lead to right-sided asymmetry in MFS of related time series.
Results are supported by the analysis of suitable surrogates shown in the corresponding figures.
Table \ref{tab:models} summarizes the main characteristics of the multifractal time series computed on all network models taken into account in our study.
\begin{table}[htp!]\scriptsize
	\centering
	\caption{Main MFS characteristics obtained for artificial networks. Only cases with pronounced multifractality are taken into account here. Relations between degree of small-worldness $S$ and right-sided asymmetry of the multifractal spectrum $A_\alpha$ is distinctly visible. Note that for SHM networks, measure $S$ (\ref{eq:smallworldness}) cannot be used due to their tree-like structure.}
	\label{tab:models}
	
	\begin{tabular}{|c|c|c|c|c|c|c||c|}\hline 
		\textbf{Model} & \multicolumn{2}{|c|}{\textbf{VCC}} & 
		\multicolumn{2}{|c|}{\textbf{VD}} 
		& \multicolumn{2}{|c|}{\textbf{VCL}} & $\boldsymbol{S}$ (\ref{eq:smallworldness}) - degree of \\
		\cline{2-3} \cline{4-5} \cline{6-7} \textbf{network} & $\Delta \alpha$ & $A_{\alpha}$ & $\Delta \alpha$ & $A_{\alpha}$ & $\Delta \alpha$ & $A_{\alpha}$ & small-worldness\\ \hline 
		WS ($k=6, p=0.01$) & 0.57  & +0.39 & - & - & 0.39 & +0.65 & 28.3 \\ \hline
		WS ($k=6, p=0.05$) & 0.30  &-0.65 & 0.14 & -0.24 & 0.23 & -0.73 & 46.2\\ \hline
		WS ($k=6, p=0.1$) & 0.21 &-0.71 & 0.15 & -0.30 & 0.12 & -0.54 & 50.3\\ \hline
		WS ($k=6, p=0.2$) & 0.17 &-0.39 & - & - & 0.10 & -0.28 & 44.1\\ \hline
		WS ($k=6, p=0.3$) & 0.15 &-0.20 & - & - & 0.13 & -0.14 & 33.14\\ \hline
		WS ($k=6, p=0.4$) & 0.15 &-0.02 & - & - & - & - & 23.8 \\ \hline
		WS ($k=4, p=0.1$) & 0.30 & -0.48 & 0.19 & -0.40 & 0.32 & -0.57 & 50.3 \\ \hline
		WS ($k=8, p=0.1$) & 0.23 & -0.30 & - & - & - & - & 41.3\\ \hline
		WS ($k=10, p=0.1$) & 0.22 & -0.18 & - & - & - & - & 33.3\\ \hline
		WS ($k=12, p=0.1$) & 0.20 & -0.15 & - & - & - & - & 26.4\\ \hline
		WS ($k=14, p=0.1$) & 0.15 & -0.20 & - & - & - & - & 28.3\\ \hline \hline
		DGM (7,8,9) & - & - & 1.03 & -0.47 & - & - & 244.1 - 1153.88\\ \hline \hline
		SHM ($e=0$) & - & - & 0.62 & -0.19 & - & - & - \\ \hline
		SHM ($e=0.1$) & 0.22 & -0.88 & 0.66 & -0.24 & - & - & - \\ \hline
		SHM ($e=0.2$) & 0.13 & -0.93 & 0.75 & -0.37 & - & - & - \\ \hline
		SHM ($e=0.3$) & 0.39 & -0.48 & 0.71 & -0.28 & - & - & - \\ \hline
		SHM ($e=0.4$) & 0.26 & -0.40 & 0.77 & -0.33 & - & - & - \\ \hline
		SHM ($e=0.5$) & - & - & 0.78 & -0.38 & - & - & - \\ \hline
		SHM ($e=0.6$) & 0.19 & -0.57 & 1.00 & -0.52 & - & - & - \\ \hline
		SHM ($e=0.7$) & 0.27 & -0.57 & 0.95 & -0.47 & - & - & - \\ \hline
		SHM ($e=0.8$) & 0.61 & -0.81 & 0.92 & -0.41 & - & - & - \\ \hline
		SHM ($e=0.9$) & 0.22 & -0.54 & 1.01 & -0.51 & - & - & - \\ \hline
		SHM ($e=1$) & 0.22 & -0.43 & 1.03 & -0.51 & - & - & - \\ \hline
	\end{tabular} 
\end{table}

%\clearpage
\subsection{Protein contact networks}

In this last experimental section, we analyze real-world data describing E. coli protein molecules \cite{mixbionets1,protgen1}. We consider network representations of folded proteins (i.e., native structures) called protein contact networks (PCNs) \cite{di2012proteins}.
Such networks are formed by considering amino acids as vertices (alpha-carbon are taken as representatives of amino acids), which are linked by edges that depend on their proximity (between 4 and 8 \AA) as given by their three-dimensional coordinates.
PCNs posses mixed topological features proper of several different prototypical models, including small-world, scale-free, fractal, and modular networks. In particular, PCNs posses low ASP and high (average) clustering coefficient, typical of small-world networks, although it is not possible to claim that PCNs are pure small-world networks (see \cite{protgen1} and references therein for details).
Here, we take into account four sample proteins with PDB codes 3DMQ (PCN0058), 4JOM (PCN0179), 2JGD (PCN0715), and 2QTA (PCN0110).

\begin{figure}[ht!]
\centering

\subfigure[MFS of PCNs.]{
\includegraphics[scale=0.3,keepaspectratio=true]{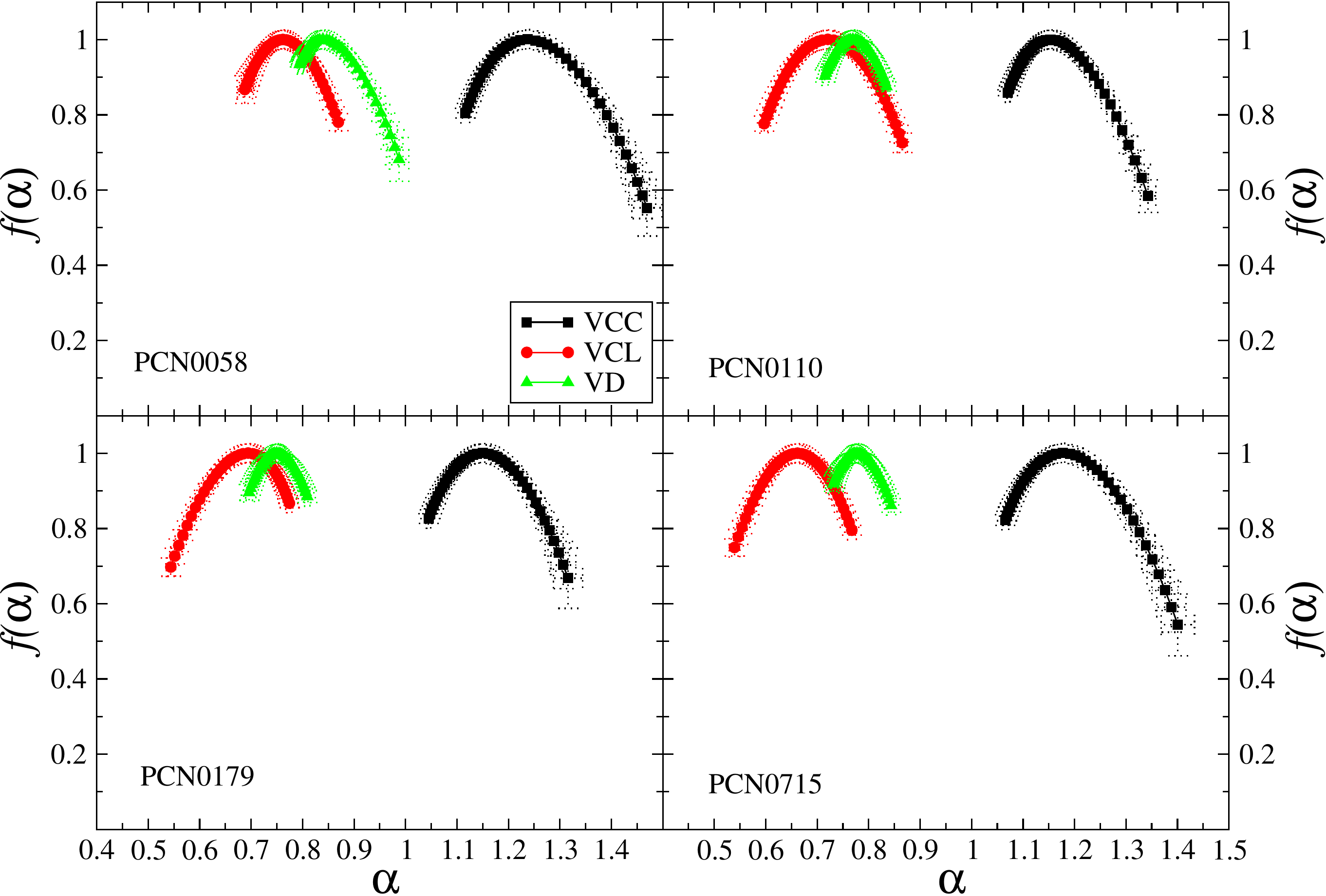}
\label{fig:PCN_MFS}}

\subfigure[Surrogates.]{
\includegraphics[scale=0.3,keepaspectratio=true]{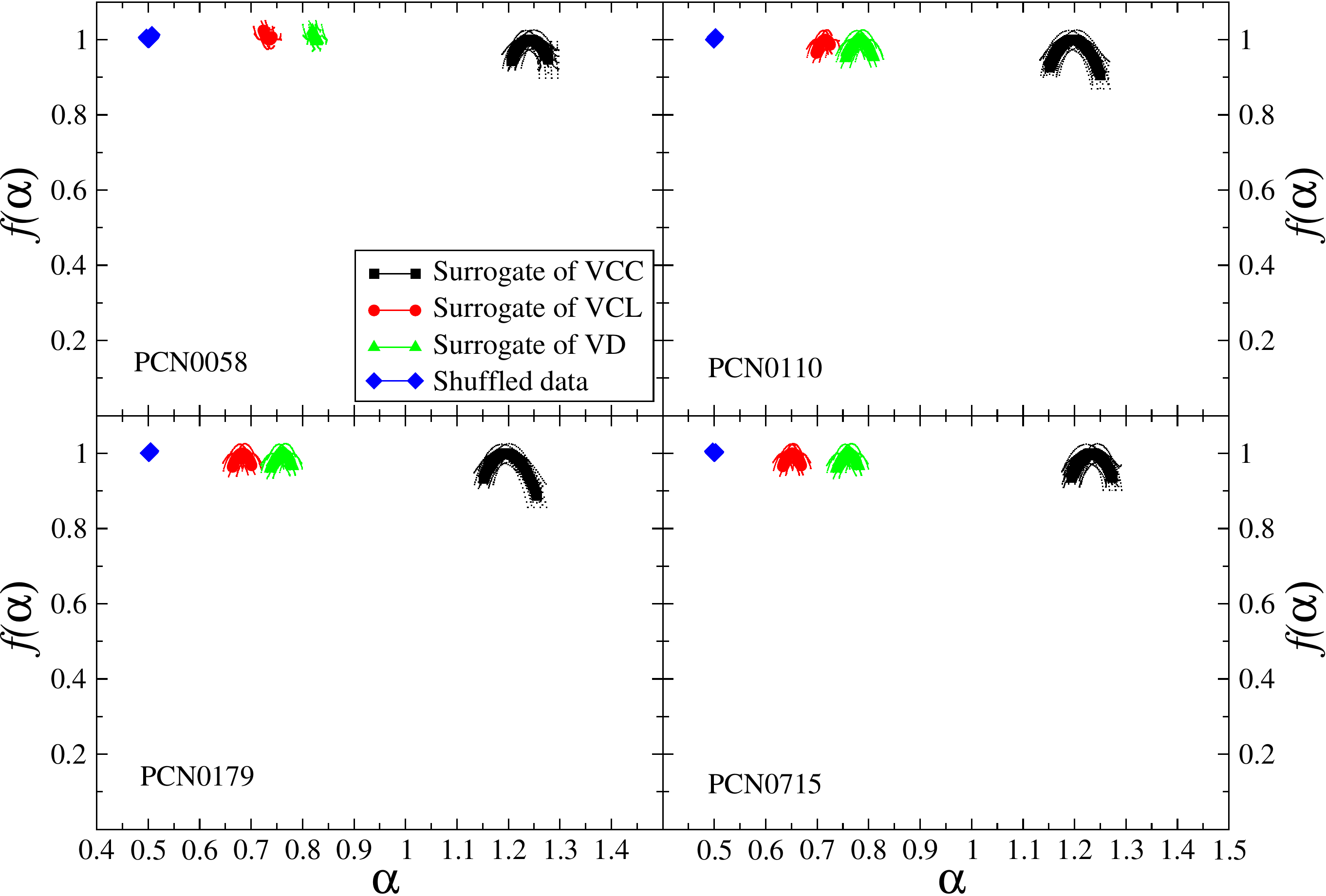}
\label{fig:PCN_surrogates}}

\caption{Results for E.coli proteins with PDB code 3DMQ (PCN0058), 4JOM (PCN0179), 2JGD (PCN0715), and 2QTA (PCN0110). Right-sided asymmetry of the MFS is clearly observed for the VCC time series.}
\label{fig:PCN}
\end{figure}

Results of multifractal analysis are shown in Fig. \ref{fig:PCN}, where we show MFS computed for the four PCNs taken into account (Fig. \ref{fig:PCN_MFS}) and related validation on surrogate time series (Fig. \ref{fig:PCN_surrogates}).
Results change depending on the vertex observable taken into account. VD time series denote a clear multifractal, right-sided spectrum only in the PCN0058 case.
VCC time series consistently denote right-sided MFS for all four PCNs.
Finally, VCL time series posses some multifractal features, although the MFS are fairly narrow; right-sided asymmetry is never observed in this case.
Table \ref{tab:pcn} provides relevant details regarding PCN topological features and calculations for the degree of small-worldness as given by Eq. \ref{eq:smallworldness} and MFS asymmetry (\ref{eq:mfs_asymmetry}) for VCC time series.
It is possible to notice that the degree of right-sided asymmetry is consistent with other topological features typically observed in small-world networks.

\begin{table}[th!]\small
\centering
\caption{Results for the four PCNs taken into account. Average values over ten random realizations are reported for the asymmetry of VCC time series.}
\label{tab:pcn}
\begin{tabular}{cccccc|cc}
\hline
\textbf{PCN} & \textbf{Vertices} & \textbf{Edge} & $\zeta$ (\ref{eq:edge_density}) & \textbf{ASP} & \textbf{ACL} & $S$ (\ref{eq:smallworldness}) & $A_{\alpha}$ (\ref{eq:mfs_asymmetry}) of \textbf{VCC} \\
\hline
PCN0058 & 938 & 3320 & 0.007 & 9.15 & 0.26 & 9.94 & -0.35 \\
PCN0179 & 909 & 3554 & 0.008 & 8.93 & 0.28 & 10.97 & -0.31 \\
PCN0715 & 811 & 3244 & 0.009 & 7.81 & 0.26 & 11.65 & -0.41 \\
PCN0110 & 801 & 3327 & 0.01 & 7.69 & 0.28 & 12.74 & -0.49 \\
\hline
\end{tabular}
\end{table}

%\clearpage
\section{Discussion and final remarks}
\label{sec:conclusions}

Networks possessing small-world features are ubiquitous in Nature and society.
Several parametric models allow to obtain networks with hybrid features, often spanning across different network types with orthogonal characteristics (e.g., fractal and small-world features, in principle, cannot co-exist within the same network).
However, considering the availability of large volume of data, it is important to design methods that allow to measure the presence of said features also in experimental networks.
Here, we addressed the important issue of assessing the degree of small-worldness in complex networks.
The proposed approach is based on fractal analysis of time series generated from networks.
Our results suggest the possibility to consider the degree of right-sided asymmetry of multifractal spectra, indicated as $A_\alpha$ in the paper, as a predictor of the degree of small-worldness present in the corresponding networks.
We validated this claim on several models, including prototypical small-world networks, scale-free, fractal and also real-world networks describing protein native structures.
The relation between the degree of small-worldness $S$ proposed in \cite{humphries2008network} and the degree of right-sided asymmetry $A_\alpha$ is consistent for Watts-Strogatz and Dorogovtsev-Goltsev-Mendes models. However, for Song-Havlin-Makse fractal models, $S$ cannot be computed, while the criterion based on $A_\alpha$ still produces consistent outcomes.
Our main result indicates that (i) $A_\alpha$ provides a reliable criterion to assess small-worldness in networks and (ii) right-sided asymmetry of multifractal spectra emerges with the presence of the following topological properties: low edge density, low average shortest path, and high clustering coefficient.
The last claim is in agreements with the findings of \citet{humphries2008network}, which showed that Watts-Strogatz small-world networks become indistinguishable from Erd\"{o}s-R\'{e}nyi graphs if, in the former, edge density is significantly increased.

%\clearpage
\bibliographystyle{abbrvnat}
\bibliography{./Bibliography.bib}

\begin{thebibliography}{47}
\providecommand{\natexlab}[1]{#1}
\providecommand{\url}[1]{\texttt{#1}}
\expandafter\ifx\csname urlstyle\endcsname\relax
  \providecommand{\doi}[1]{doi: #1}\else
  \providecommand{\doi}{doi: \begingroup \urlstyle{rm}\Url}\fi

\bibitem[Amaral et~al.(2000)Amaral, Scala, Barthelemy, and
  Stanley]{amaral2000classes}
L.~A.~N. Amaral, A.~Scala, M.~Barthelemy, and H.~E. Stanley.
\newblock Classes of small-world networks.
\newblock \emph{Proceedings of the National Academy of Sciences}, 97\penalty0
  (21):\penalty0 11149--11152, 2000.
\newblock \doi{10.1073/pnas.200327197}.

\bibitem[Avena-Koenigsberger et~al.(2015)Avena-Koenigsberger, Go{\~n}i,
  Sol{\'e}, and Sporns]{avena2015network}
A.~Avena-Koenigsberger, J.~Go{\~n}i, R.~Sol{\'e}, and O.~Sporns.
\newblock Network morphospace.
\newblock \emph{Journal of The Royal Society Interface}, 12\penalty0
  (103):\penalty0 20140881, 2015.
\newblock \doi{10.1098/rsif.2014.0881}.

\bibitem[Bassett and Bullmore(2016)]{bassett2016small}
D.~S. Bassett and E.~T. Bullmore.
\newblock Small-world brain networks revisited.
\newblock \emph{The Neuroscientist}, pages 1--18, 2016.
\newblock \doi{10.1177/1073858416667720}.

\bibitem[Bianconi(2015)]{bianconi2015interdisciplinary}
G.~Bianconi.
\newblock Interdisciplinary and physics challenges of network theory.
\newblock \emph{EPL (Europhysics Letters)}, 111\penalty0 (5):\penalty0 56001,
  2015.
\newblock \doi{10.1209/0295-5075/111/56001}.

\bibitem[Bradley and Kantz(2015)]{bradley2015nonlinear}
E.~Bradley and H.~Kantz.
\newblock Nonlinear time-series analysis revisited.
\newblock \emph{Chaos: An Interdisciplinary Journal of Nonlinear Science},
  25\penalty0 (9):\penalty0 097610, 2015.
\newblock \doi{10.1063/1.4917289}.

\bibitem[Budroni et~al.(2017)Budroni, Baronchelli, and
  Pastor-Satorras]{budroni2016scale}
M.~A. Budroni, A.~Baronchelli, and R.~Pastor-Satorras.
\newblock Scale-free networks emerging from multifractal time series.
\newblock \emph{Physical Review E}, 95:\penalty0 052311, May 2017.
\newblock \doi{10.1103/PhysRevE.95.052311}.

\bibitem[Burioni and Cassi(2005)]{burioni2005random}
R.~Burioni and D.~Cassi.
\newblock Random walks on graphs: ideas, techniques and results.
\newblock \emph{Journal of Physics A: Mathematical and General}, 38\penalty0
  (8):\penalty0 R45, 2005.
\newblock \doi{10.1088/0305-4470/38/8/R01}.

\bibitem[{Di Paola} et~al.(2012){Di Paola}, Paci, Santoni, {De Ruvo}, and
  Giuliani]{di2012proteins}
L.~{Di Paola}, P.~Paci, D.~Santoni, M.~{De Ruvo}, and A.~Giuliani.
\newblock Proteins as sponges: a statistical journey along protein structure
  organization principles.
\newblock \emph{Journal of Chemical Information and Modeling}, 52\penalty0
  (2):\penalty0 474--482, 2012.
\newblock \doi{10.1021/ci2005127}.

\bibitem[Dorogovtsev et~al.(2002)Dorogovtsev, Goltsev, and
  Mendes]{dorogovtsev2002pseudofractal}
S.~N. Dorogovtsev, A.~V. Goltsev, and J.~F.~F. Mendes.
\newblock Pseudofractal scale-free web.
\newblock \emph{Physical Review E}, 65\penalty0 (6):\penalty0 066122, 2002.
\newblock \doi{10.1103/PhysRevE.65.066122}.

\bibitem[Dro{\.z}d{\.z} and O{\'s}wi\c{e}cimka(2015)]{drozdz2015detecting}
S.~Dro{\.z}d{\.z} and P.~O{\'s}wi\c{e}cimka.
\newblock Detecting and interpreting distortions in hierarchical organization
  of complex time series.
\newblock \emph{Physical Review E}, 91\penalty0 (3):\penalty0 030902, 2015.
\newblock \doi{10.1103/PhysRevE.91.030902}.

\bibitem[Dro{\.z}d{\.z} et~al.(2009)Dro{\.z}d{\.z}, Kwapie{\'n},
  O{\'s}wiecimka, and Rak]{drozdz2009quantitative}
S.~Dro{\.z}d{\.z}, J.~Kwapie{\'n}, P.~O{\'s}wiecimka, and R.~Rak.
\newblock Quantitative features of multifractal subtleties in time series.
\newblock \emph{EPL (Europhysics Letters)}, 88\penalty0 (6):\penalty0 60003,
  2009.
\newblock \doi{10.1209/0295-5075/88/60003}.

\bibitem[Dro{\.z}d{\.z} et~al.(2016)Dro{\.z}d{\.z}, O{\'s}wi\c{e}cimka, Kulig,
  Kwapie{\'n}, Bazarnik, Grabska-Gradzi{\'n}ska, Rybicki, and
  Stanuszek]{drozdz2016}
S.~Dro{\.z}d{\.z}, P.~O{\'s}wi\c{e}cimka, A.~Kulig, J.~Kwapie{\'n},
  K.~Bazarnik, I.~Grabska-Gradzi{\'n}ska, J.~Rybicki, and M.~Stanuszek.
\newblock Quantifying origin and character of long-range correlations in
  narrative texts.
\newblock \emph{Information Sciences}, 331:\penalty0 32--44, 2016.
\newblock \doi{10.1016/j.ins.2015.10.023}.

\bibitem[Fell and Wagner(2000)]{fell2000small}
D.~A. Fell and A.~Wagner.
\newblock The small world of metabolism.
\newblock \emph{Nature Biotechnology}, 18\penalty0 (11):\penalty0 1121--1122,
  2000.
\newblock \doi{10.1038/81025}.

\bibitem[Fern{\'a}ndez-Mart{\'\i}nez et~al.(2013)Fern{\'a}ndez-Mart{\'\i}nez,
  S{\'a}nchez-Granero, and Segovia]{fernandez2013measuring}
M.~Fern{\'a}ndez-Mart{\'\i}nez, M.~A. S{\'a}nchez-Granero, and J.~E.~T.
  Segovia.
\newblock Measuring the self-similarity exponent in {L}{\'e}vy stable processes
  of financial time series.
\newblock \emph{Physica A: Statistical Mechanics and its Applications},
  392\penalty0 (21):\penalty0 5330--5345, 2013.
\newblock \doi{10.1016/j.physa.2013.06.026}.

\bibitem[Gallos et~al.(2012)Gallos, Makse, and Sigman]{gallos2012small}
L.~K. Gallos, H.~A. Makse, and M.~Sigman.
\newblock A small world of weak ties provides optimal global integration of
  self-similar modules in functional brain networks.
\newblock \emph{Proceedings of the National Academy of Sciences}, 109\penalty0
  (8):\penalty0 2825--2830, 2012.
\newblock \doi{10.1073/pnas.1106612109}.

\bibitem[Gao et~al.(2016)Gao, Small, and Kurths]{0295-5075-116-5-50001}
Z.-K. Gao, M.~Small, and J.~Kurths.
\newblock Complex network analysis of time series.
\newblock \emph{EPL (Europhysics Letters)}, 116\penalty0 (5):\penalty0 50001,
  2016.
\newblock \doi{10.1209/0295-5075/116/50001}.

\bibitem[Giuliano and Aiello(2013)]{Pagani20132688}
A.~P. Giuliano and M.~Aiello.
\newblock The power grid as a complex network: {A} survey.
\newblock \emph{Physica A: Statistical Mechanics and its Applications},
  392\penalty0 (11):\penalty0 2688--2700, 2013.
\newblock ISSN 0378-4371.
\newblock \doi{10.1016/j.physa.2013.01.023}.

\bibitem[Grech(2016)]{grech2016alternative}
D.~Grech.
\newblock Alternative measure of multifractal content and its application in
  finance.
\newblock \emph{Chaos, Solitons \& Fractals}, 2016.
\newblock \doi{10.1016/j.chaos.2016.02.017}.

\bibitem[Gu and Zhou(2010)]{PhysRevE.82.011136}
G.-F. Gu and W.-X. Zhou.
\newblock Detrending moving average algorithm for multifractals.
\newblock \emph{Physical Review E}, 82:\penalty0 011136, Jul. 2010.
\newblock \doi{10.1103/PhysRevE.82.011136}.

\bibitem[Havlin et~al.(1999)Havlin, Buldyrev, Bunde, Goldberger, Ivanov, Peng,
  and Stanley]{havlin1999scaling}
S.~Havlin, S.~V. Buldyrev, A.~Bunde, A.~L. Goldberger, {\relax P Ch}.~Ivanov,
  C.-K. Peng, and H.~E. Stanley.
\newblock Scaling in nature: from {DNA} through heartbeats to weather.
\newblock \emph{Physica A: Statistical Mechanics and its Applications},
  273\penalty0 (1):\penalty0 46--69, 1999.
\newblock \doi{10.1016/S0378-4371(99)00340-4}.

\bibitem[H{\"o}ll and Kantz(2015)]{holl2015relationship}
M.~H{\"o}ll and H.~Kantz.
\newblock The relationship between the detrendend fluctuation analysis and the
  autocorrelation function of a signal.
\newblock \emph{The European Physical Journal B}, 88\penalty0 (12):\penalty0
  1--7, 2015.
\newblock \doi{10.1140/epjb/e2015-60721-1}.

\bibitem[Humphries and Gurney(2008)]{humphries2008network}
M.~D. Humphries and K.~Gurney.
\newblock Network `small-world-ness': a quantitative method for determining
  canonical network equivalence.
\newblock \emph{PLoS ONE}, 3\penalty0 (4):\penalty0 e0002051, 2008.
\newblock \doi{10.1371/journal.pone.0002051}.

\bibitem[Kantelhardt et~al.(2002)Kantelhardt, Zschiegner, Koscielny-Bunde,
  Havlin, Bunde, and Stanley]{kantelhardt2002multifractal}
J.~W. Kantelhardt, S.~A. Zschiegner, E.~Koscielny-Bunde, S.~Havlin, A.~Bunde,
  and H.~E. Stanley.
\newblock Multifractal detrended fluctuation analysis of nonstationary time
  series.
\newblock \emph{Physica A: Statistical Mechanics and its Applications},
  316\penalty0 (1):\penalty0 87--114, 2002.
\newblock \doi{10.1016/S0378-4371(02)01383-3}.

\bibitem[Kiyono and Tsujimoto(2016)]{kiyono2016nonlinear}
K.~Kiyono and Y.~Tsujimoto.
\newblock Nonlinear filtering properties of detrended fluctuation analysis.
\newblock \emph{Physica A: Statistical Mechanics and its Applications},
  462:\penalty0 807--815, 2016.
\newblock \doi{10.1016/j.physa.2016.06.129}.

\bibitem[Kwapie{\'n} and Dro{\.z}d{\.z}(2012)]{kwapien2012physical}
J.~Kwapie{\'n} and S.~Dro{\.z}d{\.z}.
\newblock Physical approach to complex systems.
\newblock \emph{Physics Reports}, 515\penalty0 (3):\penalty0 115--226, 2012.
\newblock \doi{10.1016/j.physrep.2012.01.007}.

\bibitem[Lambiotte et~al.(2007)Lambiotte, Ausloos, and
  Ho{\l}yst]{lambiotte2007majority}
R.~Lambiotte, M.~Ausloos, and J.~A. Ho{\l}yst.
\newblock Majority model on a network with communities.
\newblock \emph{Physical Review E}, 75\penalty0 (3):\penalty0 030101, 2007.
\newblock \doi{10.1103/PhysRevE.75.030101}.

\bibitem[Livi et~al.(2016{\natexlab{a}})Livi, Maiorino, Giuliani, Rizzi, and
  Sadeghian]{protgen1}
L.~Livi, E.~Maiorino, A.~Giuliani, A.~Rizzi, and A.~Sadeghian.
\newblock A generative model for protein contact networks.
\newblock \emph{Journal of Biomolecular Structure and Dynamics}, 34\penalty0
  (7):\penalty0 1441--1454, 2016{\natexlab{a}}.
\newblock \doi{10.1080/07391102.2015.1077736}.

\bibitem[Livi et~al.(2016{\natexlab{b}})Livi, Maiorino, Pinna, Sadeghian,
  Rizzi, and Giuliani]{mixbionets1}
L.~Livi, E.~Maiorino, A.~Pinna, A.~Sadeghian, A.~Rizzi, and A.~Giuliani.
\newblock Analysis of heat kernel highlights the strongly modular and
  heat-preserving structure of proteins.
\newblock \emph{Physica A: Statistical Mechanics and its Applications},
  441:\penalty0 199--214, 2016{\natexlab{b}}.
\newblock ISSN 0378-4371.
\newblock \doi{10.1016/j.physa.2015.08.059}.

\bibitem[Maiorino et~al.(2015)Maiorino, Livi, Giuliani, Sadeghian, and
  Rizzi]{mixbionets2}
E.~Maiorino, L.~Livi, A.~Giuliani, A.~Sadeghian, and A.~Rizzi.
\newblock Multifractal characterization of protein contact networks.
\newblock \emph{Physica A: Statistical Mechanics and its Applications},
  428:\penalty0 302--313, 2015.
\newblock ISSN 0378-4371.
\newblock \doi{10.1016/j.physa.2015.02.026}.

\bibitem[Masuda et~al.(2016)Masuda, Porter, and Lambiotte]{masuda2016random}
N.~Masuda, M.~A. Porter, and R.~Lambiotte.
\newblock Random walks and diffusion on networks.
\newblock \emph{arXiv preprint arXiv:1612.03281}, 2016.

\bibitem[Mukli et~al.(2015)Mukli, Nagy, and Eke]{mukli2015multifractal}
P.~Mukli, Z.~Nagy, and A.~Eke.
\newblock Multifractal formalism by enforcing the universal behavior of scaling
  functions.
\newblock \emph{Physica A: Statistical Mechanics and its Applications},
  417:\penalty0 150--167, 2015.
\newblock \doi{10.1016/j.physa.2014.09.002}.

\bibitem[Newman(2002)]{newman2002assortative}
M.~E.~J. Newman.
\newblock Assortative mixing in networks.
\newblock \emph{Physical Review Letters}, 89\penalty0 (20):\penalty0 208701,
  2002.
\newblock \doi{10.1103/PhysRevLett.89.208701}.

\bibitem[Newman(2010)]{newman2010networks}
M.~E.~J. Newman.
\newblock \emph{Networks: An Introduction}.
\newblock Oxford University Press, Oxford, UK, 2010.

\bibitem[Nicosia et~al.(2014)Nicosia, {De Domenico}, and
  Latora]{nicosia2013characteristic}
V.~Nicosia, M.~{De Domenico}, and V.~Latora.
\newblock Characteristic exponents of complex networks.
\newblock \emph{EPL (Europhysics Letters)}, 106\penalty0 (5):\penalty0 58005,
  2014.
\newblock \doi{10.1209/0295-5075/106/58005}.

\bibitem[O\ifmmode \acute{s}\else \'{s}\fi{}wi\ifmmode~\mbox{\c{e}}\else
  \c{e}\fi{}cimka et~al.(2006)O\ifmmode \acute{s}\else
  \'{s}\fi{}wi\ifmmode~\mbox{\c{e}}\else \c{e}\fi{}cimka,
  Kwapie\ifmmode~\acute{n}\else \'{n}\fi{}, and Dro\ifmmode \dot{z}\else
  \.{z}\fi{}d\ifmmode~\dot{z}\else \.{z}\fi{}]{PhysRevE.74.016103}
P.~O\ifmmode \acute{s}\else \'{s}\fi{}wi\ifmmode~\mbox{\c{e}}\else
  \c{e}\fi{}cimka, J.~Kwapie\ifmmode~\acute{n}\else \'{n}\fi{}, and
  S.~Dro\ifmmode \dot{z}\else \.{z}\fi{}d\ifmmode~\dot{z}\else \.{z}\fi{}.
\newblock Wavelet versus detrended fluctuation analysis of multifractal
  structures.
\newblock \emph{Physical Review E}, 74:\penalty0 016103, Jul. 2006.
\newblock \doi{10.1103/PhysRevE.74.016103}.

\bibitem[O{\'s}wi\c{e}cimka et~al.(2016)O{\'s}wi\c{e}cimka, Livi, and
  Dro{\.z}d{\.z}]{mfcrosscorr_nets}
P.~O{\'s}wi\c{e}cimka, L.~Livi, and S.~Dro{\.z}d{\.z}.
\newblock Multifractal cross-correlation effects in two-variable time series of
  complex network vertex observables.
\newblock \emph{Physical Review E}, 94:\penalty0 042307, Oct. 2016.
\newblock \doi{10.1103/PhysRevE.94.042307}.

\bibitem[Riley et~al.(2012)Riley, Bonnette, Kuznetsov, Wallot, and
  Gao]{riley2012tutorial}
M.~A. Riley, S.~Bonnette, N.~Kuznetsov, S.~Wallot, and J.~Gao.
\newblock A tutorial introduction to adaptive fractal analysis.
\newblock \emph{Frontiers in Physiology}, 3, 2012.
\newblock \doi{10.3389/fphys.2012.00371}.

\bibitem[Rozenfeld et~al.(2010)Rozenfeld, Song, and Makse]{rozenfeld2010small}
H.~D. Rozenfeld, C.~Song, and H.~A. Makse.
\newblock Small-world to fractal transition in complex networks: A
  renormalization group approach.
\newblock \emph{Physical Review Letters}, 104:\penalty0 025701, Jan. 2010.
\newblock \doi{10.1103/PhysRevLett.104.025701}.

\bibitem[Schreiber and Schmitz(2000)]{schreiber2000surrogate}
T.~Schreiber and A.~Schmitz.
\newblock Surrogate time series.
\newblock \emph{Physica D: Nonlinear Phenomena}, 142\penalty0 (3):\penalty0
  346--382, 2000.
\newblock \doi{10.1016/S0167-2789(00)00043-9}.

\bibitem[Skarpalezos et~al.(2013)Skarpalezos, Kittas, Argyrakis, Cohen, and
  Havlin]{PhysRevE.88.012817}
L.~Skarpalezos, A.~Kittas, P.~Argyrakis, R.~Cohen, and S.~Havlin.
\newblock Anomalous biased diffusion in networks.
\newblock \emph{Physical Review E}, 88\penalty0 (1):\penalty0 012817, Jul.
  2013.
\newblock \doi{10.1103/PhysRevE.88.012817}.

\bibitem[Song et~al.(2006)Song, Havlin, and Makse]{song2006origins}
C.~Song, S.~Havlin, and H.~A. Makse.
\newblock Origins of fractality in the growth of complex networks.
\newblock \emph{Nature Physics}, 2\penalty0 (4):\penalty0 275--281, 2006.
\newblock \doi{10.1038/nphys266}.

\bibitem[Wang et~al.(2016)Wang, Lai, and Grebogi]{wang2016data}
W.-X. Wang, Y.-C. Lai, and C.~Grebogi.
\newblock Data based identification and prediction of nonlinear and complex
  dynamical systems.
\newblock \emph{Physics Reports}, 644:\penalty0 1--76, 2016.
\newblock \doi{10.1016/j.physrep.2016.06.004}.

\bibitem[Watts and Strogatz(1998)]{watts1998collective}
D.~J. Watts and S.~H. Strogatz.
\newblock Collective dynamics of 'small-world' networks.
\newblock \emph{Nature}, 393\penalty0 (6684):\penalty0 440--442, 1998.
\newblock \doi{10.1038/30918}.

\bibitem[Wendt and Abry(2007)]{wendt2007multifractality}
H.~Wendt and P.~Abry.
\newblock Multifractality tests using bootstrapped wavelet leaders.
\newblock \emph{IEEE Transactions on Signal Processing}, 55\penalty0
  (10):\penalty0 4811--4820, 2007.
\newblock \doi{10.1109/TSP.2007.896269}.

\bibitem[Weng et~al.(2014)Weng, Zhao, Small, and Huang]{weng2014time}
T.~Weng, Y.~Zhao, M.~Small, and D.~D. Huang.
\newblock Time-series analysis of networks: {E}xploring the structure with
  random walks.
\newblock \emph{Physical Review E}, 90\penalty0 (2):\penalty0 022804, 2014.
\newblock \doi{10.1103/PhysRevE.90.022804}.

\bibitem[Weng et~al.(2017)Weng, Zhang, Small, Zheng, and Hui]{weng2017memory}
T.~Weng, J.~Zhang, M.~Small, R.~Zheng, and P.~Hui.
\newblock Memory and betweenness preference in temporal networks induced from
  time series.
\newblock \emph{Scientific Reports}, 7:\penalty0 41951, 2017.
\newblock \doi{10.1038/srep41951}.

\bibitem[Zhou(2009)]{zhou2009components}
W.-X. Zhou.
\newblock The components of empirical multifractality in financial returns.
\newblock \emph{EPL (Europhysics Letters)}, 88\penalty0 (2):\penalty0 28004,
  2009.
\newblock \doi{10.1209/0295-5075/88/28004}.

\end{thebibliography}
\end{document}